\documentclass[11pt]{article} 
\usepackage{amsfonts,amsmath,latexsym,amssymb,mathrsfs,amsthm,comment,setspace}
\usepackage{slashbox, bm, natbib}
\usepackage{caption}
\usepackage{enumitem}
\usepackage{url, placeins, xr}
\usepackage{algorithmic, float, graphicx}

\let\OLDthebibliography\thebibliography
\renewcommand\thebibliography[1]{
  \OLDthebibliography{#1}
  \setlength{\parskip}{0pt}
  \setlength{\itemsep}{0pt plus 0.0ex}
}

\newcommand{\Prob}{{\rm pr}}

\newcommand{\bea}{\begin{eqnarray}}
\newcommand{\ena}{\end{eqnarray}}
\newcommand{\beq}{\begin{equation}}
\newcommand{\enq}{\end{equation}}
\newcommand{\beas}{\begin{eqnarray*}}
\newcommand{\enas}{\end{eqnarray*}}

\def\numberlikeadb{\global\def\theequation{\thesection.\arabic{equation}}}
\numberlikeadb
\newtheorem{theorem}{Theorem}[section]

\usepackage{color} 

\usepackage{lscape}
\usepackage{caption}
\usepackage{multirow}
\RequirePackage{amsthm,amsmath,amsfonts,amssymb,color}
\RequirePackage{graphicx}
\usepackage{tikz}
\usepackage{xr}
\usepackage{lscape}
\usepackage{longtable}

\begin{document}

\title{Generalized multiple change-point detection in the structure of multivariate, possibly high-dimensional, data sequences}
\author{Andreas Anastasiou\footnote{Department of Mathematics and Statistics, University of Cyprus, P.O. Box: 20537, 1678, Nicosia, Cyprus, 
anastasiou.andreas@ucy.ac.cy}\:\: and Angelos Papanastasiou\footnote{Department of Mathematics and Statistics, University of Cyprus, P.O. Box: 20537, 1678, Nicosia, Cyprus, papanastasiou.angelos@ucy.ac.cy}}

\date{} 
\maketitle

\vspace{-10mm}


\begin{abstract}
The extensive emergence of big data techniques has led to an increasing interest in the development of change-point detection algorithms that can perform well in a multivariate, possibly high-dimensional setting. In the current paper, we propose a new method for the consistent estimation of the number and location of multiple generalized change-points in multivariate, possibly high-dimensional, noisy data sequences. The number of change-points is allowed to increase with the sample size and the dimensionality of the given data sequence. Having a number of univariate signals, which constitute the unknown multivariate signal, our algorithm can deal with general structural changes; we focus on changes in the mean vector of a multivariate piecewise-constant signal, as well as changes in the linear trend of any of the univariate component signals. Our proposed algorithm, labeled Multivariate Isolate-Detect (MID), allows for consistent change-point detection in the presence of frequent changes of possibly small magnitudes in a computationally fast way.
\end{abstract}

\noindent{{\bf{Keywords:}}} Change-point detection; high-dimensional setting; piecewise-constant structure; piecewise-linear structure.

\section{Introduction}

\label{sec:intro}
Change-point detection algorithms have been actively developed and investigated from the scientific community. Their ability to segment data into smaller, homogeneous parts have helped researchers and practitioners to develop flexible statistical models that can adapt in non-stationary environments. Due to the natural data heterogeneity in many real problems, such algorithms have been applied in a wide range of application areas, such as bioinformatics (\cite{picard2011joint, hocking2013learning}), cyber security (\cite{siris2004application}), or finance (\cite{lavielle2007adaptive,schroder2013adaptive}). The advantages of detecting changes in the behaviour of the data fall into two main categories; interpretation and forecasting. Interpretation comes naturally since the detected changes are usually connected with life events that took place near the estimation time. Associating the changes with such real-life phenomena can lead to a better understanding and quantification of the effect that these events had on the behaviour of the stochastic process. With respect to forecasting, the role of the final segment is important because it allows for a more accurate prediction of the future values of the data sequence at hand.

Based on whether we have full knowledge of the data to be analysed, change-point detection is split into two main categories; offline detection, where the data are already obtained, and online detection, in which the observations arrive sequentially at present. With respect to the dimensionality of the data, change-point detection can be further separated into algorithms that act only on univariate data and to those that are suitable for change-point detection in multivariate data sequences. In this paper, we focus on multivariate, possibly high-dimensional, offline settings; our aim is to estimate the number and locations of certain structural changes in the behaviour of given multivariate data. The model is
\begin{equation}\label{eq:model}
    \boldsymbol{X_{t}} = \boldsymbol{f_{t}} + \Sigma\boldsymbol{\epsilon_{t}}, \hspace{1cm} t=1, \ldots, T,
\end{equation}
where $\boldsymbol{X_{t}} \in \mathbb{R}^{d\times 1}$ are the observed data and $\boldsymbol{f_{t}} \in \mathbb{R}^{d \times 1}$ is the $d$-dimensional deterministic signal with structural changes at certain points. The signals that we treat in the current manuscript are those that changes appear in the mean structure or in the vector of the first order derivatives. The diagonal matrix $\Sigma \in \mathbb{R}^{d \times d}$ has diagonal elements denoted by $\sigma_1, \ldots, \sigma_d$, while for any $t \in \left\lbrace 1,\ldots, T\right\rbrace$, the noise terms $\boldsymbol{\epsilon_{t}} \in \mathbb{R}^{d \times 1}$ are random vectors with mean the zero vector and covariance the identity matrix. The elements $\sigma_j, j = 1,\ldots, d$ of the diagonal matrix $\Sigma$ might be unknown. In such cases, $\sigma_j$ are estimated using the Median Absolute Deviation method explained in \cite{Hampel}. The true number, $N$, of the change-points, as well as their locations $r_1, \ldots, r_N$, are unknown and our aim is to estimate them; $N$ is free to grow with the sample size $T$ and the dimensionality $d$.

The initial purpose of change-point detection algorithms has been to detect a single change in the mean structure of a univariate signal under the setting of Gaussian noise, but much progress has since been made. Researchers have heavily focused on the detection of multiple change-points in the mean structure of a univariate data sequence. Towards this purpose, optimization-based methods have been developed, in which the estimated signal is chosen based on its fit to the data, penalized by a complexity rule. To solve the implied penalization problem, dynamic programming approaches, such as the Segment Neighborhood (SN) and Optimal Partitioning (OP) methods of \cite{auger_lawrence} and \cite{jackson}, have been developed. \cite{killick2012optimal} and \cite{rigaill} introduce improvements over the classical OP and SN algorithms, respectively. In the context of regression problems, \cite{frick2014multiscale} introduced the simultaneous multiscale change-point estimator (SMUCE) for change-point detection in exponential family regression. Apart from optimization based algorithms, a popular method in the literature is binary segmentation where changes are detected one at a time through an iterative binary splitting of the data. Recent variants of binary segmentation with improved performance are the Wild Binary Segmentation (WBS) of \cite{fryzlewicz2014wild} and its recently developed second version of \cite{fryzlewicz2020wild}, the Narrowest-Over-Threshold (NOT) method of \cite{NOT}, and the Seeded Binary Segmentation of \cite{seeded}. The Isolate-Detect (ID) algorithm has been developed in \cite{anastasiou2019detecting} to detect, one by one, structural changes in a data sequence. It is based on an isolation technique, which leads to very good accuracy on the estimated number and locations of the change-points, particularly in scenarios with many frequent change-points. Our proposed method is partly based on ID; therefore we elaborate on its important parts later. For a more thorough review of the literature on the detection of multiple change-points in the mean of univariate data sequences, see \cite{Cho_Kirch_2020} and \cite{Yu2020}. Apart from changes in the mean of a univariate data sequence, research has also been done for the detection of change-points under more complex scenarios, such as detection of changes in the slope for piecewise-linear models (\cite{anastasiou2019detecting, NOT, CPOP, Maeng_Fryzlewicz, Tibshirani}) changes in the variance (\cite{Inclan_Tiao}), as well as for distributional changes under a non-parametric setting (\cite{matteson, Zouetal2014}).

Even though there is an extensive literature on change-point detection for univariate data sequences, the multivariate, possibly high-dimensional setting, which is the focus of this paper, has not been investigated in such degree. Working under the model in \eqref{eq:model}, \cite{vert2010fast} proposes a method for approximating the signal $\boldsymbol{f_{t}}$ as the solution of a convex optimization problem. In order to achieve this, the problem is first reformulated to a group LASSO one and then the group least-angle regression (LARS) procedure explained in \cite{yuan2006model} is employed. Another interesting approach with very good behaviour has been introduced in \cite{wang2016high}. The algorithm, called {\textit{inspect}}, estimates the number and locations of the change-points in the mean structure of $\boldsymbol{f_t}$ as in \eqref{eq:model}. Firstly, {\textit{inspect}} applies a cumulative sum (CUSUM) transformation to the original data matrix. Secondly, a projection direction of the transformed matrix is computed as its leading left singular vector, and, finally, a univariate change-point detection algorithm is applied to the projected series. It is among the many methods that employ CUSUM-type statistics for change-point detection in the multivariate setting. In general, methods that belong to this category, mainly either use CUSUM aggregations of the $d$ component data sequences in order to test the obtained values against a threshold or to construct alternative test-statistics. For instance, \cite{groen2013multivariate} examines the asymptotic behaviour of the maximum absolute and average CUSUM and gives finite-sample performance results. Focusing on testing for the existence of a change-point in the mean structure of the multivariate signal, \cite{enikeeva2013high} and \cite{horvath2012change} propose tests based on the $\ell_2$ aggregation of the CUSUM statistics for each univariate component, while \cite{jirak2015uniform} employs the $\ell_\infty$ aggregation of the aforementioned values. In \cite{cho2016change}, the Double CUSUM (DC) operator is introduced, which takes as input the ordered absolute CUSUM values of each individual component and performs a weighted $\ell_1$ aggregation to construct the DC statistic which is then compared against a test criterion. Departing from the detection of changes in the mean structure of a multivariate signal, \cite{cho2015multiple} propose the Sparsified Binary Segmentation algorithm (SBS) for the detection of multiple change-points in the second-order structure of a multivariate data sequence. SBS is based on a first, ``sparsifying'' step which is used to exclude individual component data sequences from an $\ell_1$ aggregation; a pre-defined threshold is used for the exclusion. The recent work of \cite{CCID} introduces Cross-covariance isolate detect (CCID), which, motivated from the necessity of estimating changes in time-varying functional connectivity networks, detects multiple change-points in the second-order (cross-covariance or network) structure of multivariate, possibly high-dimensional time series. \cite{ombao2005slex} investigates the application of smooth localized complex exponentials (SLEX) waveforms, to the detection of changes in spectral characteristics of EEG data. \cite{lavielle2006detection} detects changes in the covariance structure of i.i.d. multivariate time series based on the minimization of a penalized Gaussian log likelihood, while \cite{bucher2014detecting} uses a test statistic based on sequential empirical copula processes to detect changes in the cross-covariance structure. For a survey of various offline change-point detection algorithms on multivariate time series see \cite{truong2020selective}.\par

In this paper, we propose a method called Multivariate Isolate-Detect (MID) for the consistent estimation of multiple change-points under the multivariate, possibly high-dimensional, structure of the model in \eqref{eq:model}. Our method builds on the foundations of the ID algorithm developed in \cite{anastasiou2019detecting}; that is, we first isolate each true change-point within subintervals of the domain $[1,\ldots,T]$ and then we proceed to detect them. Isolation enhances detection power, especially in frameworks with frequently occurring change-points. MID is explained in detail in Section \ref{sec:methodology and theory}; here, we only give a brief description of its important steps. The main idea is that for the observed data sequences $x_{t,j} \hspace{0.2cm} t=1,\ldots, T,\quad j=1,\ldots,d,$ and with $\lambda_{T}$ a positive constant, playing the role of an expansion step as in \cite{anastasiou2019detecting}, our method first creates two ordered sets of $K=\lceil T/\lambda_{T}\rceil$ right- and left-expanding intervals. For $i =1,\ldots, K$, the $i^{th}$ right expanding interval is $R_i=[1,\min\left\lbrace i\lambda_T, T\right\rbrace]$ while the $i^{th}$ left-expanding interval is $L_{i}=[\max\left\lbrace 1, T-i\lambda_{T}+1\right\rbrace,T].$ We collect these intervals in the ordered set $S_{RL}=\{R_1,L_1,R_2,L_2,\ldots,R_K,L_K\}$. The algorithm first acts on the interval $R_1 = [1, \lambda_T]$ by calculating, for every univariate component data sequence, the contrast function value for the $Q$ possible candidates in this interval (details are given in Section \ref{sec:methodology and theory}). This process will return $Q$ vectors $\boldsymbol{y_j}, j=1,\ldots,Q$ of length $d$ each; for example, the elements of $\boldsymbol{y_1} \in \mathbb{R}^{d}$ will be the contrast function values related to the first change-point candidate in $R_1$, for each of the $d$ component data sequences, the elements of $\boldsymbol{y_2} \in \mathbb{R}^{d}$ will be the relevant values for the second candidate in $R_1$, and so on. The next step is to apply to each $\boldsymbol{y_j}$ a mean-dominant norm $L: \mathbb{R}^d \rightarrow \mathbb{R}$. The definition of such norms can be found in Section 2 of \cite{Carlstein} and examples include
\begin{align}
\label{mean_dominant}
\nonumber & L_2 := L_2(\boldsymbol{y_j}) = \frac{1}{\sqrt{d}}\sqrt{\sum_{i=1}^d y_{j,i}^2}\\
& L_{\infty} := L_{\infty}(\boldsymbol{y_j}) = \sup_{i=1,\ldots,d}\left\lbrace y_{j,i}\right\rbrace,
\end{align}
where $y_{j,i} \geq 0, \;\forall i, j$. Applying $L(\cdot)$ to each $\boldsymbol{y_j}$, will return a vector $\boldsymbol{v}$ of length $Q$. We identify $\tilde{b}_{R_1}$ := ${\rm argmax}_j\left\lbrace v_j \right\rbrace$. If $v_{\tilde{b}_{R_1}}$ exceeds a certain threshold, denoted by $\zeta_{T,d}$, then $\tilde{b}_{R_1}$ is taken as a change-point. If not, then the process tests the next interval in $S_{RL}$. Upon detection, the algorithm makes a new start from the end-point (respectively, start-point) of the right- (respectively, left-) expanding interval where the detection occurred. Upon correct choice of $\zeta_{T,d}$, MID ensures that we work on intervals with at most one change-point.

The rest of the paper is structured as follows. Section \ref{sec:methodology and theory} describes our proposed methodology and its associated theory. Useful variants of our algorithm and the choice of important parameter values are explained in Section \ref{sec:Computational}. In Section \ref{sec:simulations}, we perform a thorough simulation study to compare our algorithm with state-of-the-art methods, while Section \ref{sec:Real_data} illustrates the behaviour of MID on two examples of real data; the monthly percentage changes in the UK house price index over a period of twenty two years in twenty London Boroughs, and the daily number of new COVID-19 cases in the four constituent countries of the United Kingdom; England, Northern Ireland, Scotland, and Wales. We conclude the paper in Section \ref{sec:conclusion} with some general remarks and reflections on our proposed algorithm. The proofs of Theorem \ref{theorem_consistency_S1} and Theorem \ref{theorem_consistency_S2} are given in the Appendix.

\section{Methodology and Theory}\label{sec:methodology and theory}
\subsection{Methodology}\label{methodology}
\begin{figure*}
\includegraphics[width=0.9\textwidth, height=0.25\textheight]{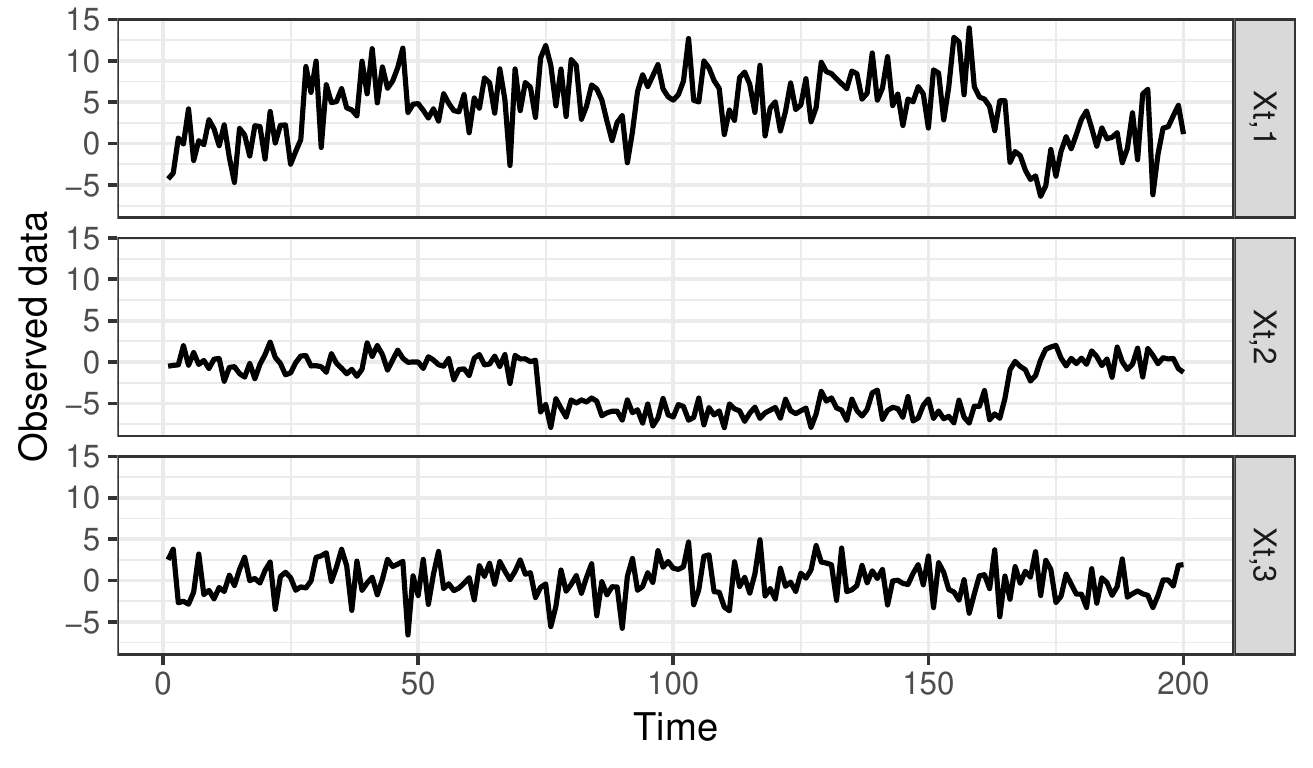}%
  \caption{An example of a three dimensional data sequence that undergoes three changes in its mean structure at locations $r_1 = 27, r_2 = 73$ and $r_3 = 165$.}
  \label{example_3d}
\end{figure*}

\begin{figure*}
  \makebox[\textwidth][c]{\includegraphics[width=0.95\textwidth, height = 0.4\textheight]{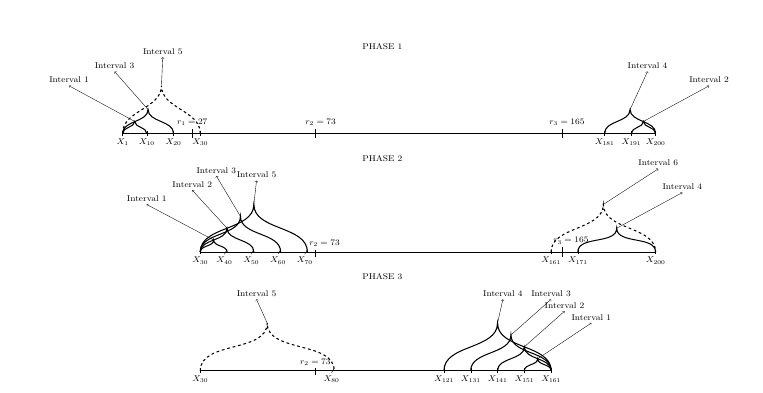}}%
  \caption{An example with three change-points in the mean structure; $r_1=27, r_2 = 73$ and $r_3 = 165$. The dashed line is the interval in which the detection took place in each phase.}
  \label{fig:steps}
\end{figure*}

\begin{figure*}
  \makebox[\textwidth][c]{\includegraphics[width=0.95\textwidth, height = 0.35\textheight]{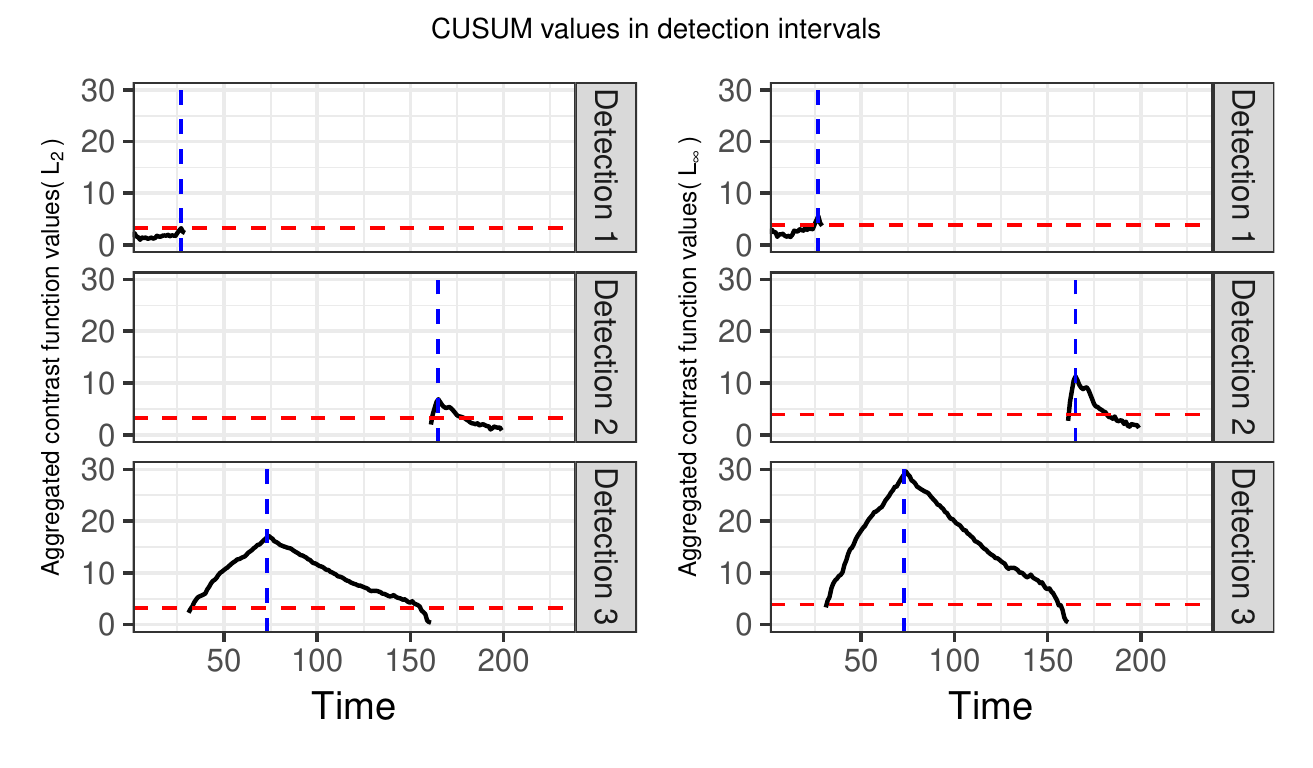}}%
  \caption{CUSUM values in the relevant detection intervals explained in Figure \ref{fig:steps}. The vertical dashed line indicates the time point with the highest value in the corresponding interval, while the horizontal dashed line is the optimal threshold value. On the left column you can see the results when the $L_2$ aggregation method was used, while for the right column we employed the $L_{\infty}$-based aggregation approach.}
  \label{cusum_step}
\end{figure*}

We work under the model given in \eqref{eq:model}. The objective is to estimate both the number, $N$, and the locations $r_1,\ldots, r_N$ where the multivariate deterministic signal $\boldsymbol{f_{t}}$ exhibits structural changes. We note that $N$ can possibly grow with the sample size $T$ and with the dimensionality $d$. In addition, a change-point does not necessarily appear in all univariate component signals. Before providing a full, step-by-step explanation of our algorithm, two simple examples are given to assist in ease of understanding. In Figure \ref{example_3d}, we have a three dimensional data sequence of length $T = 200$ with three change-points in the mean vector at locations $r_1=27, r_2=73$ and $r_3=165$. To be more precise,
\begin{align}
\label{example}
\nonumber & f_{t,1} = \begin{cases}
    0, & t=1,\ldots,27\\
    6, & t=28,\ldots,165\\
    0, & t=166,\ldots,200
  \end{cases},\\
& f_{t,2} = \begin{cases}
    0, & t=1,\ldots,73\\
    -6, & t=74,\ldots,165\\
    0, & t=166,\ldots,200
  \end{cases},\\ 
\nonumber & f_{t,3} = 0, t = 1, \ldots, 200.
\end{align}
We take $\sigma_1 = 3, \sigma_2 = 1,$ and $\sigma_3 = 2$, while the random variables $\epsilon_{t,i}$ follow the standard Gaussian distribution for $i=1,2,3$. The component data sequences $X_{t,1}$ and $X_{t,2}$ share a common change-point at $t=165$, while they also have their own change-points at $t=27$ and $t=73$, respectively. There are no change-points in $X_{t,3}$.
Let us denote by $r_0=0$, $r_{N+1}=T$. In this toy example, we take the expansion parameter $\lambda_T=10$, while $\zeta_{T,d}$ is a well-chosen predefined threshold. The choice of the aforementioned parameters is discussed in detail in Section \ref{subsec:parameter_choice}; special attention will be given to the dimensionality of the data sequence in order to make a robust threshold choice. Figure \ref{fig:steps} shows the steps of MID until all change-points are detected. We will be referring to Phases 1, 2 and 3 involving five, six, and five intervals, respectively; these are clearly indicated in Figure \ref{fig:steps}. These phases are only related to this specific example of three change-points; in cases with a different number of change-points we would have a different number of such phases. At the beginning of the detection process, we have that $s = 1$ and $e = T = 200$. As already mentioned in Section \ref{sec:intro}, the proposed algorithm acts both sequentially and interchangeably on subintervals of the full domain; this being $[1,\ldots,200]$ for this example. For a well-chosen threshold $\zeta_{T,d}$, then $r_1=27$ is the first change-point to be detected; this occurs in the interval $[1,30]$ as shown in Phase 1 of Figure \ref{fig:steps}. We now briefly explain how the detection occurred for $r_1$. Let $A$ be the $30 \times 3$ matrix, with each column being the first 30 observations (since we are working in the interval $[1,30]$) of each of the three univariate data sequences. The next step is to compute the contrast function (in this specific case of piecewise-constant signals, the function is the absolute value of the widely used CUSUM statistic as given in \eqref{cusum_formula}) values for each candidate point and for all three component data sequences. We end up with a matrix $B \in \mathbb{R}^{29 \times 3}$, with $B_{i,j}$ being the value of the contrast function for the $i^{th}$ data point of the $j^{th}$ data sequence when we work in the interval $[1,30]$; the last point of the interval is not among the change-point candidates. Applying a mean-dominant norm to each row of $B$ gives us a vector of length 29. Figure \ref{cusum_step} (Detection 1), illustrates these values, when we employed the $L_2$ and the $L_\infty$ mean-dominant norms as defined in \eqref{mean_dominant}.
In Figure \ref{cusum_step}, we see that for both employed norms, $t=27$ has the highest value, which exceeds the predefined threshold value obtained as in Section \ref{subsec:parameter_choice}. Therefore, $\hat{r}_1 = 27$ is assigned as the estimated location for $r_1$. After the detection, $s$ is updated as the end-point of the (right-expanding) interval where the detection occurred; therefore $s = 30$, and MID is, in Phase 2, applied in the interval $[30,200]$. Then, $r_3=165$ gets detected at the sixth step of Phase 2 in the interval $[161,200]$. After this second detection, MID proceeds to Phase 3, where it is applied in the interval $[30,161]$ and $r_2$ gets isolated (for the first time) and detected in the interval $[30,80]$ as shown in Figures \ref{fig:steps} and \ref{cusum_step}. In the end, MID is applied in the interval $[80,161]$, where there will be no expanding interval that contains a point with an aggregated CUSUM value that surpasses the threshold $\zeta_{T,d}$; therefore, the process will terminate after scanning all the data.

\begin{figure*}
  \makebox[\textwidth][c]{\includegraphics[width=0.9\textwidth, height = 0.3\textheight]{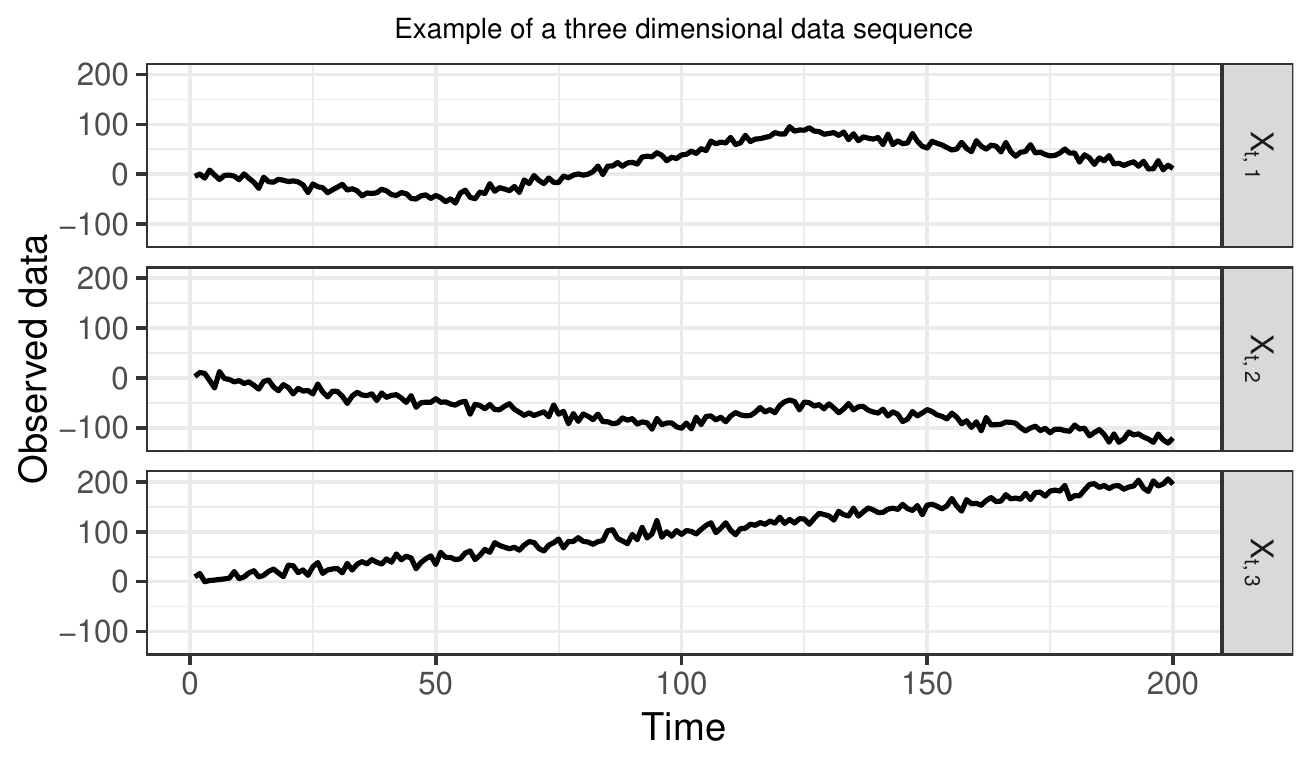}}%
  \caption{An example of a three dimensional data sequence with  piecewise-linear structure, that undergoes three changes in its first derivative at locations $r_1 = 53, r_2 = 100$ and $r_3 = 124$.}
  \label{example_linear}
\end{figure*}

In Figure \ref{example_linear}, we graphically provide an example of a three dimensional data sequence of length $T = 200$ with three change-points in the slope at locations $r_1=53, r_2=100$ and $r_3=124$. To be more precise,
\begin{align}
\nonumber
& f_{t,1} = \begin{cases}
    -t+1, & t=1,\ldots,53\\
    2t-158, & t=54,\ldots,124\\
    -t+214, & t=125,\ldots,200
  \end{cases}, \\
\nonumber & f_{t,2} = \begin{cases}
-t+1, & t=1,\ldots,100\\
    2t-299, & t=101,\ldots,124\\
    -t+73, & t=125,\ldots,200
  \end{cases},\\
\nonumber & f_{t,3} = t, t = 1, \ldots, 200.
\end{align}
We take $\sigma_1 = \sigma_2 = \sigma_3 = 7$, while $\epsilon_{t,i}$ follow the standard Gaussian distribution for $i=1,2,3$. The first two component data sequences have two change-points each; for $X_{t,1}$ at locations $t=53$ and $t=124$, while for $X_{t,2}$ at locations $t = 100$ and $t=124$. There are no change-points in $X_{t,3}$.
After giving two examples of structures that MID can be employed to, our method can now be described in a general framework. Our proposed algorithm is based on the same isolation technique as that of the univariate change-point detection method ID and therefore, extensive details of how this isolation is achieved are avoided and can be found in Section 3.1 of \cite{anastasiou2019detecting}. As shown in Figure \ref{fig:steps} for the specific example in Figure \ref{example_3d}, our method is looking for change-points interchangeably in right- and left-expanding intervals. In each of them, MID acts in the same way: Let $I$ be one of these intervals, with cardinality $\mid I\mid $, and $A^I \in \mathbb{R}^{\mid I\mid  \times d}$ is such that $A^I_{i,j} = X_{i,j}$ for $i \in I$ and $j \in \left\lbrace 1,\ldots,d\right\rbrace$. The contrast function value is then separately applied to the elements of each of the columns of $A^I$; this will result in the matrix $B^{I} \in \mathbb{R}^{J \times d}$, whose element at position $(i,j)$ corresponds to the contrast function value of the $i^{th}$ data point in $I$ for the $j^{th}$ data sequence. The number $J$ of rows of the matrix $B^{I}$ (with the contrast function values) depends on the type of changes that our algorithm aims to find; for example, it has already been discussed in the literature that in the case of changes in the mean structure $J = \mid I\mid -1$, while if we are looking for changes in the slope of a piecewise-linear signal, then $J = \mid I\mid -2$.

The next step is to apply to each row of $B^{I}$, a mean-dominant norm as those in \eqref{mean_dominant} which will lead to a vector $v \in \mathbb{R}^{J}$. We identify $\hat{b}_{I}={\rm argmax}_t\{v_t\}$ and $v_{\hat{b}_{I}}$ is tested against the predefined threshold $\zeta_{T,d}$; heuristically, $v_{\hat{b}_{I}}$ will be small if $b$ is not a change-point and large in the opposite scenario. If $v_{\hat{b}_{I}} > \zeta_{T,d}$, then $\hat{b}_I$ is taken as a change-point and MID restarts and looks for change-points in non-previously tested intervals. The algorithm stops when there are no more intervals to be tested.
\subsection{Theory}
\label{subsec:theory}
We work under the setting in \eqref{eq:model}. Denoting by $r_0 = 0$ and $r_{N+1} = T$, the illustration scenarios are:
\\
\textbf{(S1) Changes in the mean structure:} For $k = 1,\ldots, N+1$, $\boldsymbol{f_t} = \boldsymbol{\mu_k} \in \mathbb{R}^{d}$ for $t = r_{k-1}+1, \ldots, r_k$. In this case, the univariate component signals $f_{t,j}$, for $j = 1,\ldots, d$ are piecewise-constant.
\\
\textbf{(S2) Changes in the first derivative:} For $k = 1,\ldots, N+1$, $\boldsymbol{f_t} = \boldsymbol{\mu_{1,k}} + \boldsymbol{\mu_{2,k}}t$ for $t = r_{k-1}+1,\ldots, r_k$, where $\boldsymbol{\mu_{1,k}}$ and $\boldsymbol{\mu_{2,k}}$ are vectors in $\mathbb{R}^{d}$. In addition, we require that for $j = 1,\ldots, N$, the equality $\boldsymbol{\mu_{1,j}} + \boldsymbol{\mu_{2,j}}r_j = \boldsymbol{\mu_{1,j+1}} + \boldsymbol{\mu_{2,j+1}}r_j$ is satisfied. Under this framework, the change-points, $r_k$, satisfy that $\boldsymbol{f_{r_{k} - 1}} + \boldsymbol{f_{r_{k} + 1}} \neq 2\boldsymbol{f_{r_{k}}}$. Therefore, the univariate component signals $f_{t,j}$, for $j = 1,\ldots, d$ are continuous and piecewise-linear.

The aforementioned scenarios are only two specific illustration cases in which the proposed MID algorithm can be applied. Due to its change-point isolation step prior to detection, our algorithm can be applied in more complicated scenarios where each univariate signal could be, for example, piecewise-polynomial or piecewise-exponential. In Sections \ref{subsec:pcm} and \ref{subsec:cplm}, we provide the main theorems for the consistency of our method in accurately estimating the true number and the location of the change-points in Scenarios (S1) and (S2), respectively. The theoretical results presented in this section are for the $L_{\infty}$ norm being employed for the aggregation of the information from the component data sequences; in practice, our algorithm exhibits very good behaviour for other mean-dominant norms as well, something which is shown in Sections \ref{sec:simulations} and \ref{sec:Real_data}.
\subsubsection{Scenario (S1)}
\label{subsec:pcm}
As already discussed, the first step of the detection process depends on an appropriately chosen contrast function, which, for every component data sequence, is applied to each change-point candidate. In Scenario (S1) , the contrast function applied to the component data sequences $X_{t,j}, \forall j \in \left\lbrace 1,\ldots, d\right\rbrace$ is the absolute value of the widely used CUSUM statistic, with the latter being defined as
\begin{equation}
\label{cusum_formula}
\tilde{X}_{s,e}^{b,j} = \sqrt{\frac{e-b}{n(b-s+1)}}\sum_{t=s}^{b}X_{t,j} - \sqrt{\frac{b-s+1}{n(e-b)}}\sum_{t=b+1}^{e}X_{t,j},
\end{equation}
where $1\leq s \leq b < e\leq T$ and $n=e-s+1$. Before proceeding with the main theoretical result for the consistency of our method, allow us to introduce some more notation, as below.
\begin{align}
\label{general_notation}
\nonumber & \delta_T := \min_{j=1,\ldots, N+1}\mid r_j - r_{j-1} \mid, \\
\nonumber & \Delta_{j}^{q} := \mid f_{r_{j+1},q} - f_{r_j,q} \mid, \quad j=1,\ldots, N,\quad q = 1, \ldots, d\\
& \underline{f}_{T} := \inf_{j=1,\ldots,N}\left\lbrace\sup_{q=1,\ldots,d}\Delta_{j}^{q}\right\rbrace.
\end{align}
For the consistency result with respect to the estimated number of change-points and their estimated locations obtained by MID, we work under the assumption (A1) as follows:
\begin{itemize}
{\small{\item[(A1)] The minimum distance, $\delta_T$, between two change-points and the minimum magnitude of jumps, $\underline{f}_T$ as in \eqref{general_notation}, are connected by $\sqrt{\delta_T}\underline{f}_T \geq \underline{C}\sqrt{\log \left(Td^{1/4}\right)}$, for a large enough constant $\underline{C}$.}}
\end{itemize}
The number of change-points, $N$, is allowed to grow with the sample size $T$ and the dimensionality $d$. Theorem \ref{theorem_consistency_S1} provides the main theoretical result for Scenario (S1). The proof is given in Appendix \ref{sec:Proof_S1}.
\begin{theorem}
\label{theorem_consistency_S1}
Let $\left\lbrace \boldsymbol{X_t} \right\rbrace_{t=1,\ldots,T}$ follow model \eqref{eq:model} with $\boldsymbol{f_t}$ as in Scenario (S1) and $\boldsymbol{\epsilon_{t}}$ are i.i.d. random vectors from the $d$-variate standard normal distribution. Let $N$ and $r_j, j=1,\ldots,N$ be the number and locations of the change-points, while $\hat{N}$ and $\hat{r}_j, j=1,\ldots,\hat{N}$ are their estimates sorted in increasing order. In addition, for $[s_j, e_j]$ being the interval where $\hat{r}_j$ is obtained, we denote by $q_j:= {\rm argmax}_{k=1,\ldots,d}\mid \tilde{X}_{s_j,e_j}^{\hat{r}_j,k} \mid$. Assuming that (A1) holds, then, there exist positive constants $C_1, C_2, C_3,$ and $C_4$, which do not depend on $T$ or $d$, such that for $C_1\sqrt{\log \left(Td^{1/4}\right)} \leq \zeta_{T,d} < C_2\sqrt{\delta_T}\underline{f}_T$, 
\begin{equation}
\label{mainresult_theorem}
\nonumber \Prob\left(V \leq C_3\log\left(Td^{\frac{1}{4}}\right)\right) \geq 1 - \frac{C_4}{T},
\end{equation}
where $V = \left\lbrace \hat{N} = N, \underset{j = 1,\ldots, N}{\max}\left\lbrace \mid \hat{r}_j - r_j \mid \left(\Delta_j^{q_j}\right)^2\right\rbrace \right\rbrace$.
\end{theorem}
The lower bound for the probability in \eqref{mainresult_theorem} does not depend on the dimensionality $d$ and its order is $1 - \mathcal{O}\left(T^{-1}\right)$. Furthermore, the rate of convergence of the estimated change-point locations does not depend on the minimum distance between two change-points, $\delta_T$; the jump magnitude, $\Delta_j^{q_j}$, is the only quantity that affects the rate. We notice, though, that to be able to match the estimated change-point locations with the true ones, then $\delta_T$ should be larger than the distance between the estimated and the true change-point locations. Therefore, based on \eqref{mainresult_theorem}, we deduce that $\delta_T$ must be at least $\mathcal{O}\left(\log \left(Td^{1/4}\right)\right)$. This, combined with Assumption (A1) that requires $\delta_T(\underline{f}_T)^2$ to be of order at least $\mathcal{O}\left(\log \left(Td^{1/4}\right)\right)$, means that $\underline{f}_T$ could decrease with $T$ in cases where $\delta_T$ is of order higher than $\mathcal{O}\left(\log \left(Td^{1/4}\right)\right)$.

With respect to the threshold parameter, $\zeta_{T,d}$, the rate of its lower bound is $\mathcal{O}\left(\sqrt{\log \left(Td^{1/4}\right)}\right)$; this will also be used in practice as the default rate. Therefore, we have that
\begin{equation}
\label{threshold}
\zeta_{T,d} = C\sqrt{\log \left(Td^{1/4}\right)},
\end{equation}
where $C$ is a positive constant. More details on the choice of $C$ are given in Section \ref{subsec:parameter_choice}.
\subsubsection{Scenario (S2)}
\label{subsec:cplm}
We are under the scenario where for any $j \in \left\lbrace 1,\ldots, d\right\rbrace$, the underlying signal $f_{t,j}, t=1,\ldots,T$ has a continuous and piecewise-linear structure as in Figure \ref{example_linear}. In this case, the contrast function applied to the component data sequences $X_{t,j}, \forall j \in \left\lbrace 1,\ldots, d\right\rbrace$ is
\begin{equation}
\label{contrast_linear}
C_{s,e}^b(\boldsymbol{X_j}) = \mid \left\langle\boldsymbol{X_j},\boldsymbol{\phi_{s,e}^b}\right\rangle \mid,
\end{equation}
where for $n=e-s+1$ and
\begin{align}
\nonumber & \alpha_{s,e}^b = \sqrt{\frac{6}{n(n^2-1)(1+(e-b+1)(b-s+1)+(e-b)(b-s))}}\\
\nonumber & \beta_{s,e}^b = \sqrt{\frac{(e-b+1)(e-b)}{(b-s+1)(b-s)}},
\end{align}
we have that the contrast vector, $\phi_{s,e}^b(t)$, is equal to
\begin{align}
\label{contrast_vectorCPLM}
\nonumber & \alpha_{s,e}^b\beta_{s,e}^b\left[(e+2b-3s+2)t - (be +bs - 2s^2+2s)\right], \text{ for } t \in [s,b]\\
\nonumber & \frac{\alpha_{s,e}^b}{\beta_{s,e}^b}\left[(2e^2+2e-be-bs)-(3e-2b-s+2)t\right], \text{ for } t \in [b+1,e]\\
& 0, \text{otherwise}.
\end{align}
For more details on how this vector is constructed for (S2), please see section B.2 in the online supplementary material of \cite{NOT}. Proceeding now with the consistency result of MID applied to Scenario (S2), we make the following assumption.
\begin{itemize}
{\small{\item[(A2)] The minimum distance, $\delta_T$, between two change-points and the minimum magnitude of changes, $\underline{f}_T:= \underset{j=1,\ldots,N}{\inf}\left\lbrace\underset{q=1,\ldots,d}{\sup} \mid f_{r_{j}-1,q} + f_{r_{j}+1,q} - 2f_{r_{j},q} \mid \right\rbrace$, are connected by $\delta_T^{3/2}\underline{f}_T \geq \underline{C}^*\sqrt{\log \left(Td^{1/4}\right)}$, where $\underline{C}^*$ is a large enough constant.}}
\end{itemize}
Theorem \ref{theorem_consistency_S2} provides the main theoretical result for Scenario (S2). The proof can be found in Appendix \ref{app:proofs}.
\begin{theorem}
\label{theorem_consistency_S2}
Let $\left\lbrace \boldsymbol{X_t} \right\rbrace_{t=1,\ldots,T}$ follow model \eqref{eq:model} with $\boldsymbol{f_t}$ as in Scenario (S2) and $\boldsymbol{\epsilon_{t}}$ are i.i.d. random vectors from the $d$-variate standard normal distribution. Let $N$ and $r_j, j=1,\ldots,N$ be the number and locations of the change-points, while $\hat{N}$ and $\hat{r}_j, j=1,\ldots,\hat{N}$ are their estimates sorted in increasing order. In addition, for $[s_j, e_j]$ denoting the interval where $\hat{r}_j$ is obtained during MID, we denote by $q_j:= {\rm argmax}_{k=1,\ldots,d}\left\lbrace C_{s_j,e_j}^{\hat{r}_j}(\boldsymbol{X_k})\right\rbrace$. With $\Delta_j^q := \mid f_{r_{j}-1,q} + f_{r_{j}+1,q} - 2f_{r_{j},q} \mid$ and assuming that (A2) holds, then, there exist positive constants $C_1, C_2, C_3,$ and $C_4$, which do not depend on $T$ or $d$, such that for $C_1\sqrt{\log \left(Td^{1/4}\right)} \leq \zeta_{T,d} < C_2\delta_T^{3/2}\underline{f}_T$, we obtain that
\begin{equation}
\label{mainresult_theorem_S2}
\Prob\left(\tilde{V} \leq C_3\left(\log\left(Td^{\frac{1}{4}}\right)\right)^{1/3}\right) \geq 1 - \frac{C_4}{T},
\end{equation}
where $\tilde{V} = \left\lbrace \hat{N} = N, \underset{j = 1,\ldots, N}{\max}\left\lbrace\mid\hat{r}_j - r_j\mid\left(\Delta_j^{q_j}\right)^{2/3}\right\rbrace\right\rbrace$.
\end{theorem}
Similarly as in (S1), the upper bound for the probability in \eqref{mainresult_theorem_S2} does not depend on the dimensionality of the data sequence $\boldsymbol{X_t}$; its rate of convergence is $1 - \mathcal{O}\left(T^{-1}\right)$. Furthermore, the rate of convergence of the estimated change-point locations depends only on the change magnitude, $\Delta_j^{q_j}$, as defined in the statement of Theorem \ref{theorem_consistency_S2}. The lower bound for the threshold, $\zeta_{T,d}$, is of order $\mathcal{O}\left(\sqrt{\log \left(Td^{1/4}\right)}\right)$; this is the default rate that is used in practise in Sections \ref{sec:simulations} and \ref{sec:Real_data}. Therefore,
\begin{equation}
\label{threshold_S2}
\zeta_{T,d} = C^*\sqrt{\log \left(Td^{1/4}\right)},
\end{equation}
where $C^* > 0$. More details on the choice of the values for $C^*$ are given in Section \ref{subsec:parameter_choice}.

\section{Practicalities and variants}
\label{sec:Computational}
\subsection{Mean-dominant norms}
The proposed MID methodology is based on an aggregation step of the contrast function values obtained from each component data sequence using mean-dominant norms in the way these are defined in Section 2 of \cite{Carlstein}. Even though Theorems \ref{theorem_consistency_S1} and \ref{theorem_consistency_S2} cover the theoretical behaviour of MID under the $L_{\infty}$ norm given in \eqref{mean_dominant}, similar results could be obtained when other norms are employed. In the remaining sections, the discussion on the choice of the parameter values as well as results on the practical performance of MID will be focused when our method is combined with the $L_{\infty}$ or the $L_2$ mean-dominant norm, as defined in \eqref{mean_dominant}, for the aggregation step of the contrast function values.
\subsection{Choice of parameter values}
\label{subsec:parameter_choice}
In order to choose the constants $C$ in \eqref{threshold} and $C^*$ in \eqref{threshold_S2}, we ran a large scale simulation study involving data sequences $\left\lbrace \boldsymbol{X_t} \right\rbrace_{t=1,\ldots,T}$, for $T = 700, 1400$ following model \eqref{eq:model}, where $\boldsymbol{f_t}=\boldsymbol{0}$ while $\boldsymbol{\epsilon_{t}}$ are i.i.d. from the $d$-variate standard normal distribution. Specifically, for each $d \in \{1,\ldots,50\}$ we generated 500 replicates and applied MID under scenarios (S1) and (S2) to each one of those replicates using various threshold constant values $C$ and $C^*$, respectively, in order to estimate the number of change-points. For each dimension, in order to control the Type I error rate, $\alpha$, of falsely detecting change-points, we chose the default constant to be the one that its number of times that successfully did not detect any change-points was closer to $(1-\alpha)*500$. For $d>50$, we keep the threshold that gave the best results for the simulated 50-dimensional data sequence. Table \ref{tab:best_thres} presents the results for $\alpha \in \left\lbrace 0.05, 0.1\right\rbrace$ under Scenarios (S1) and (S2). From now on, the obtained values for $C$ and $C^*$ will be referred to as the default constants.
\begin{table*}
\centering
\caption{The optimal values for the threshold constants, $C$ and $C^*$, which control the Type I error rate $\alpha$ under the Scenarios (S1) and (S2), respectively, for $d = 1,\ldots,50$. Results are presented for the $L_2$ and the $L_\infty$ norms}
\begin{tabular}{@{}cllllll@{}}
 & \multicolumn{3}{|c|}{} & \multicolumn{3}{|c|}{}\\
 & \multicolumn{3}{|c|}{Results for the (S1) scenario} & \multicolumn{3}{|c|}{Results for the (S2) scenario} \\
Norm & $d$ & $C$ ($\alpha = 0.05$) & $C$ ($\alpha = 0.1$) & $d$ & $C^*$ ($\alpha = 0.05$) & $C^*$ ($\alpha = 0.1$)\\
\hline
& 1 & 1.7 & 1.55 & 1 & 1.65 & 1.55\\
& 2 & 1.25 & 1.25 & 2 & 1.25 & 1.2\\
& 3 & 1.1 & 1.05 & 3 & 1.05 & 1.05\\
& 4 & 1.05 & 0.95 & 4 & 0.95 & 0.95\\
& 5 & 0.95 & 0.9 & 5 & 0.9 & 0.9\\
& 6 & 0.9 & 0.9 & 6 & 0.9 & 0.85\\
$L_2$ & 7 & 0.9 & 0.8 & 7 & 0.8 & 0.8\\
& 8 & 0.8 & 0.8 & 8 & 0.8 & 0.75\\
& 9 & 0.8 & 0.75 & 9-11 & 0.75 & 0.75\\
& 10-13 & 0.75 & 0.75 & 12-16 & 0.7 & 0.7\\
& 14 & 0.75 & 0.65 & 17-19 & 0.65 & 0.6\\
& 15-20 & 0.7 & 0.65 & 20-22 & 0.6 & 0.6\\ 
& 21-23 & 0.65 & 0.6 & 24-42 & 0.6 & 0.55\\
& 24-39 & 0.6 & 0.6 &43-50& 0.55 & 0.55\\
& 40-50 & 0.6 & 0.55 &  & & \\
\hline
& 1 & 1.7 & 1.55 & 1 & 1.65 & 1.55\\
& 2-3 & 1.75 & 1.7 & 2 & 1.7 & 1.6\\
& 4-6 & 1.8 & 1.7 & 3 & 1.75 & 1.6\\
$L_{\infty}$ & 7-13 & 1.85 & 1.75 & 4-5 & 1.75 & 1.65\\
& 14-25 & 1.9 & 1.8 & 6-13 & 1.75 & 1.7\\
& 26-28 & 1.9 & 1.85 & 14-25 & 1.8 & 1.75\\
& 29-50 & 1.95 & 1.85 & 26-38 & 1.85 & 1.8\\
 & & & & 39-50 & 1.9 & 1.85\\
\end{tabular}
\label{tab:best_thres}
\end{table*}
\subsection{Decision on the aggregation method}
\label{subsec:reduction}
In \cite{cho2015multiple}, \cite{cho2016change}, and \cite{CCID}, it has been explained that the $L_\infty$-based aggregation of the contrast functions for each component data sequence tends to exhibit a better behaviour (compared to the $L_1$ and $L_2$-based aggregations) in scenarios where the true change-points appear only in a small number of the component data sequences. In contrast, due to spuriously large contrast function values, the $L_{\infty}$ approach could suffer in situations where the change-points are not sparse across the panel of the data sequences.

This difference in the behaviour between the $L_2$ and the $L_{\infty}$ norms examined in our paper is what has motivated us to introduce an introductory step in MID, where the sparsity in a given $d$-dimensional data sequence, $\boldsymbol{X_t}$, is first estimated. For $r_1, \ldots, r_N$ being the $N$ true change-points, allow us, for any $j \in \left\lbrace 1,\ldots,N\right\rbrace$, to define by $A_j \subseteq \left\lbrace 1,\ldots,d\right\rbrace$ the set of indices for the univariate component data sequences that contain the change-point $r_j$. Then, for $\boldsymbol{X_t}$, the sparsity is given by
\begin{equation}
\label{eq:sparsity}
sp = \max_{j = 1,\ldots,N}\left\lbrace\mid A_j\mid \right\rbrace/d,
\end{equation}
where $\mid A_j\mid $ is the cardinality of the set $A_j$. It is straightforward that $sp \in [0,1]$. Our proposed hybrid approach aims to data adaptively decide on the aggregation method to be used and eliminates the necessity of choosing between the $L_2$ and $L_\infty$ norms. The steps to achieve our purpose are as follows.
\\
{\textbf{Step 1:}} We apply MID paired with the $L_\infty$ aggregation rule in order to obtain $\hat{r}_1, \ldots, \hat{r}_M$. It has already been explained that the $L_\infty$ norm is preferable and provides very good results in cases with sparse change-points, while it tends to overestimate the number of change-points (due to the contrast function taking spuriously large values) under the scenario of dense change-points. Therefore, $M \geq N$.\\
{\textbf{Step 2:}} We will now estimate the sparsity in the given data as defined in \eqref{eq:sparsity}. With $\hat{r}_0 = 0$ and $\hat{r}_{M+1} = T$, we first collect the triplets $(\hat{r}_{m-1}+1,\hat{r}_m,\hat{r}_{m+1}), \forall m \in \left\lbrace 1,\ldots, M \right\rbrace$. After this, $\forall i \in \left\lbrace 1,\ldots,d \right\rbrace$, we calculate $CS^{(i)}(\hat{r}_m) := C^{\hat{r}_m}_{\hat{r}_{m-1}+1,\hat{r}_{m+1}}(\boldsymbol{X_i})$, with $C^b_{s,e}(\boldsymbol{X_i})$ being the relevant contrast function (based on whether we are under Scenario (S1) of Section \ref{subsec:pcm} or Scenario (S2) of Section \ref{subsec:cplm}) value for the point $b$, when we are in the interval $[s,e]$, for the univariate component data sequence $\boldsymbol{X_i}$. The $d$ contrast function values for each $\hat{r}_m$ are collected in the sets
\begin{equation}
\label{eq:cusum_sets}
S_m = \left\lbrace CS^{(1)}(\hat{r}_m), \ldots, CS^{(d)}(\hat{r}_m)\right\rbrace, \quad j = 1,\ldots, M.
\end{equation}
{\textbf{Step 3:}} For each $m = 1, \ldots, M$, all the elements of $S_m$ are tested against the relevant threshold value, $\zeta_T$, for univariate change-point detection which is of the order $\mathcal{O}(\sqrt{\log T})$ in both scenarios of piecewise-constant and continuous piecewise-linear signals (representing scenarios (S1) and (S2), respectively, covered in this paper). In terms of the threshold constants, we employ those of \cite{anastasiou2019detecting}. We denote by $\hat{sp}_m := \frac{\#\left\lbrace i:CS^{(i)}(\hat{r}_m) > \zeta_T\right\rbrace}{d}$. The estimated sparsity is then
$\hat{sp} = \max_{m \in \left\lbrace 1, \ldots, M\right\rbrace}\left\lbrace \hat{sp}_m \right\rbrace.$\\
{\textbf{Step 4:}} For cases where $\hat{sp} \leq 0.4$, we accept the result from Step 1, where MID has been paired with the $L_{\infty}$ norm, whereas if $\hat{sp} \geq 0.6$, then MID is paired with the $L_2$ norm. Extensive simulations have shown that there is no significant difference on the MID's practical performance with respect to accuracy (on both the estimated number of change-points and the estimated locations) when $\hat{sp} \in (0.4,0.6)$. Therefore, MID could be paired with any of the aforementioned two norms and give very good results. For computational cost reasons, in such cases we accept the result of Step 1. From now on, we denote by ${\rm MID}_{\rm opt}$ to be the data-adaptive, sparsity-based MID version explained in this section, where an aggregation method is chosen based on the estimated sparsity of the change-points in the given data.
\subsection{A permutation-based approach}
\label{subsec:permutation}
MID is a threshold-based algorithm because at each step, the largest aggregated contrast function value is tested against a predefined threshold in order to decide whether there is a change-point at the corresponding location. In Section \ref{subsec:parameter_choice} we have explained how the threshold constant is carefully chosen to control the Type I error taking into account the dimensionality of the data. However, misspecification of the threshold can possibly lead to the misestimation of the number of change-points. To solve such issues, we propose a variant of MID based on permutation.

The idea of a ``data-adaptive threshold'' through permutations or bootstrapping is not new. In \cite{cho2016change}, a bootstrap procedure is proposed, which is motivated by the representation theory developed for the Generalised Dynamic Factor Model, in order to approximate the quantiles of the developed double CUSUM test statistic under the null hypothesis of no change-points; the obtained quantiles are used as a test criterion for detection. In \cite{cabrieto2018testing}, a permutation-based approach is used to test the presence of correlation changes in multivariate time series. Under a univariate framework, in \cite{antoch2001permutation} a permutation scheme is proposed for deriving critical values for test statistics based on functionals of the partial sums $\sum_{i=1}^{k}(X_i-\bar{X}_n), k=1,\ldots,n$, where $X_i$ are the observed data and $\bar{X}_n$ their mean value. The proposed scheme consists of three steps: a) compute the test statistic using the original data, b) construct the permutation distribution by computing the relevant test statistic on permuted versions of the data, and c) reject the null hypothesis of no change-points if the test statistic lies in the tails of the distribution.

We propose a variant that combines the isolation technique of MID with an extension of the permutation procedure used in \cite{antoch2001permutation} to the multivariate framework. Although this permutation procedure tends to be computationally expensive, it has a straightforward implementation. As generally holds for permutation procedures, the test-statistic obtained from the original data is compared to those obtained when applying the same steps to several permuted versions of the data. For our proposed permutation scheme all steps remain the same as MID with the only difference being the way the algorithm chooses to accept or reject a change-point within each interval. To be more precise, suppose that, for a given data sequence $\left\lbrace \boldsymbol{X_t}\right\rbrace_{t=1,\ldots,T}$, the MID algorithm is at a step where it looks for a change-point in the interval $I=[s^*,e^*]$, where $1 \leq s^* < e^* \leq T$. As described in Section \ref{methodology}, MID returns a vector $\boldsymbol{v} \in \mathbb{R}^{J}$, where $J$ is the amount of all change-point candidates. The elements of $\boldsymbol{v}$ correspond to the aggregated contrast function values for each candidate point in $I$. The next step is to store $T_{I_{max}} = \underset{i \in \{1,\ldots, J\}}{\max}\left\lbrace v_i\right\rbrace$ and repeat the following procedure a prespecified amount, denoted by $K$, of times:
\begin{enumerate}
{\small{
    \item Generate a random permutation from $(s^{*},\ldots, e^{*})$.
    \item Reorder the data according to the permutation.
    \item Calculate and store the maximum aggregated contrast function value for each permutation.}}
\end{enumerate}
The empirical distribution obtained from the maximum values is used to construct our test. More precisely, we identify $\hat{b}_{I}={\rm argmax}_t\{v_t\}$ as a change-point if, for given $\alpha \in (0,1)$, $T_{I_{max}} > q_{1-\alpha}$, where $q_{1-\alpha}$ is the $100(1-\alpha)\%$ quantile of the empirical distribution.

The parameter $\alpha$ controls the probability of false detections. Small values of $\alpha$ will make it harder for the algorithm to reject the null hypothesis of no change-points, whereas large values can reduce the probability of a Type II error. For the simulations in Section \ref{sec:simulations}, we take $\alpha= 0.01$. Regarding the parameter $K$, the simulation results provided in \cite{antoch2001permutation}--although for a univariate framework--suggest that the empirical quantiles get stabilised quickly. Therefore, for our simulations, we take $K=1000$.
\section{Simulations}
\label{sec:simulations}
In this section, we investigate the performance of our method in Scenarios (S1) and (S2) covered in Sections \ref{subsec:pcm} and \ref{subsec:cplm}, respectively. Furthermore, in (S1) MID is compared with state-of-the-art methods through a comprehensive simulation study. The competitors are the Double Cusum (DC) method of \cite{cho2016change}, the Sparsified Binary Segmentation (SBS) algorithm of \cite{cho2015multiple}, and the INSPECT algorithm of \cite{wang2016high}. DC and SBS are implemented in the \textit{hdbinseg} \textsf{R} package, while for INSPECT we have used the \textit{InspectChangepoint} \textsf{R} package. For DC, the parameters used were the ones that gave the best results in the simulation study carried out in the relevant paper (\cite{cho2016change}), while SBS and INSPECT were called with their default arguments. Regarding our algorithm, we give results for the data-adaptive, sparsity-based MID version of Section \ref{subsec:reduction}, denoted by ${\rm MID}_{\rm opt}$, as well as for the permutation-based variants in Section \ref{subsec:permutation}; these are denoted by ${\rm MIDPERL}_2$ and ${\rm MIDPERL}_\infty$ for the $L_2$ and $L_\infty$ norms, respectively.
\begin{table*}
\centering
\caption{Distribution of $\hat{N} - N$ over 100 simulated multivariate data sequences with 3 change-points under Scenario (S1) of Section \ref{subsec:pcm}. The average ARI, $d_H$, and computational times are also given}
\begin{tabular}{@{}llllllllllll@{}}
 & & & \multicolumn{6}{|c|}{} & & & \\
Method & $sp$ & $d$ & \multicolumn{6}{|c|}{$\hat{N} - N$} & ARI & $d_h$ & Time (s)\\
 & & &$-2$ & $-1$ & 0 & 1 & 2 & 3 & & & \\
$\boldsymbol{{\rm MID}_{\rm opt}}$& 0.2 & 30 & 0 &   0 &  \textbf{95} &  5& 0 & 0 & 1 
& 0.01 & 0.43\\ 
DC & 0.2 & 30 &0 &   0 &  90 &  10 & 0 & 0 & 0.98 
& 0.03 & 5.91\\ 
SBS & 0.2 & 30 & 1 &  36 &  62 &   1  & 0 & 0 & 0.96 
& 0.15 & 1.08\\ 
\textbf{INSPECT} & 0.2 & 30 & 0 &   0 &  \textbf{96} &   4& 0 & 0 & 1 
& 0.01 & 0.31\\ 
${\rm MIDPERL}_2$ & 0.2 & 30 &  0 & 2 & 82 & 14 & 0 & 2 & 0.96 & 0.05 & 330.81 \\
${\rm MIDPERL}_{\infty}$ & 0.2 & 30 & 0 & 0 & 79 & 18 & 2 & 1 & 0.96 & 0.05 & 172.72 \\
\hline
$\boldsymbol{{\rm MID}_{\rm opt}}$& 0.2 & 100 & 0 &   1 &  \textbf{90} & 8 & 1 & 0& 1 
& 0.03 & 1.41\\ 
{\bf{DC}} & 0.2 & 100 & 0 &   1 &  \textbf{94} &   5  & 0 & 0 & 0.99 
& 0.01 & 13.11 \\ 
SBS & 0.2 & 100 & 5 &  25 &  70 &   0 & 0 & 0 & 0.97 
& 0.14 & 2.31\\ 
{\bf{INSPECT}} & 0.2 & 100 & 0 &   2 &  \textbf{90} & 7 & 1 & 0 & 1 
& 0.02 & 1.38\\ 
${\rm MIDPERL}_2$ & 0.2 & 100 &  0 &4 & 81 & 13 & 2 & 0& 0.97  & 0.04 & 1098.40  \\
${\rm MIDPERL}_{\infty}$ & 0.2 & 100 & 0 & 1 & 79 & 16 & 4 & 0 &0.95 & 0.07 & 615.85\\
\hline
$\boldsymbol{{\rm MID}_{\rm opt}}$ & 0.5 & 30 & 0 & 1 & {\bf{93}} & 6 & 0 & 0 & 1  
& 0.02 & 0.43 \\ 
{\bf{DC}} & 0.5 & 30 & 0 & 0 & {\bf{92}} & 7 & 1 & 0 & 0.99 
& 0.01 & 5.94 \\ 
SBS & 0.5 & 30 & 2 & 23 & 75 & 0 & 0 &0& 0.98 
& 0.10 & 1.08 \\ 
\textbf{INSPECT} & 0.5 & 30 & 0 & 1 & \textbf{90} & 9 & 0 & 0 & 1 
& 0.01 & 0.38\\ 
${\rm MIDPERL}_2$ & 0.5 & 30 & 0 & 2 & 80 & 15 & 2 & 1 & 0.97 &0.06 & 331.91  \\
${\rm MIDPERL}_{\infty}$ & 0.5 & 30 & 0 & 1 & 74 & 20 & 5 & 0 & 0.96 &0.05 & 161.49 \\
\hline
${\rm MID}_{\rm opt}$ & 0.5 & 100 & 0 & 0 & 83 & 16 & 1 & 0 & 0.98 
& 0.02 & 1.38 \\ 
{\bf{DC}} & 0.5 & 100 & 0 & 0 & {\bf{90}} & 9 & 1 &0 & 0.99 
& 0.02 & 13.12\\ 
SBS & 0.5 & 100 & 0 & 27 & 73 & 0 & 0 &0& 0.98 
& 0.10 &2.31 \\ 
{\bf{INSPECT}} & 0.5 & 100 & 0 & 0 & {\bf{93}} & 7 & 0 &0& 0.99
& 0.02 &1.31 \\ 
$\boldsymbol{{\rm MIDPERL}_2}$ & 0.5 & 100 & 0 & 1 & {\bf{88}} & 10 & 1 &0 & 0.98 & 0.03& 1069.54 \\
$\boldsymbol{{\rm MIDPERL}_{\infty}}$ & 0.5 & 100 & 0 & 0 & {\bf{88}} & 10 & 2 & 0 & 0.97 &0.04 & 751.34\\ 
\hline
$\boldsymbol{{\rm MID}_{\rm opt}}$ & 0.8 & 30 & 0 & 1 & \textbf{97} & 2 & 0 & 0 & 1 
& 0.01 & 0.44\\ 
DC & 0.8 & 30 & 0 & 1 & 90 & 9 & 0 & 0& 0.99 
& 0.02 & 5.94 \\ 
SBS & 0.8 & 30 & 0 & 32 & 68 & 0 & 0 & 0& 0.98 
& 0.09 & 1.08 \\ 
\textbf{INSPECT} & 0.8 & 30 & 0 & 1 & \textbf{94} & 4 & 1 & 0 &1 
& 0.01 & 0.31\\ 
${\rm MIDPERL}_2$ & 0.8 & 30 & 0 & 1 & 89 & 6 & 4 & 0 & 0.98 & 0.04 & 324.60 \\
${\rm MIDPERL}_{\infty}$ & 0.8 & 30 & 0 & 0 & 80 & 17 & 2 & 1 & 0.97& 0.07  & 161.49 \\ 
\hline
$\boldsymbol{{\rm MID}_{\rm opt}}$ & 0.8 & 100 & 0 & 2 & {\bf{97}} & 1 & 0 &0 & 1 
& 0.01 & 1.70 \\ 
{\bf{DC}} & 0.8 & 100 & 0 & 0 & {\bf{93}} & 7 & 0 &0& 0.99 
& 0.01 &13.37 \\ 
SBS & 0.8 & 100 & 4 & 24 & 72 & 0 & 0 & 0&0.97  
& 0.14 & 2.35  \\ 
\textbf{INSPECT} & 0.8 & 100 & 0 & 0 & \textbf{93} & 7 & 0 & 0 & 0.99 
& 0.01 &1.30 \\ 
${\boldsymbol{{\rm MIDPERL}_2}}$ & 0.8 & 100 & 0 & 1 & {\bf{93}} & 4 & 1 & 1 &0.98  & 0.02 &1085.59 \\
${\rm MIDPERL}_{\infty}$ & 0.8 & 100 & 0 & 0 & 85 & 12 & 3 &0 & 0.98 & 0.03 & 602.87 \\
\end{tabular}
\label{3changepointssparisty0.8}
\end{table*}
\vspace{0.01in}
\\
\textbf{Simulation setup:} We considered the settings where $T=1500$, $d \in \left\lbrace 30,100\right\rbrace$ and $N \in \left\lbrace 3,20,50 \right\rbrace$. With the definition of sparsity as in \eqref{eq:sparsity}, we took $sp \in \left\lbrace 0.2,0.5,0.8 \right\rbrace$. In total, we tested the methods in 18 different setups covering a wide range of multivariate sequences. The change magnitude at the component data sequences (either in the mean or in the slope depending on whether we are under (S1) or (S2), respectively) follows the uniform distribution ${\rm U}(1,2)$. Standard Gaussian noise was added to the signals. We ran 100 replications for each setup. The frequency distribution of $\hat{N}-N$ is provided, while the accuracy of the estimated locations is evaluated through the Adjusted Rand Index (ARI) of the estimated segmentation against the true one (\cite{hubert1985comparing}), and the scaled Hausdorff distance, 
\begin{equation}
\nonumber d_H=n_s^{-1}\max\bigg\{\max_j\min_k \mid r_j - \hat{r_k}\mid,\max_k\min_k \mid r_j-\hat{r_k} \mid,\bigg\},
\end{equation}
where $n_s$ is the length of the largest segment. The average computational times, in seconds, are also provided. The results, for MID and the competing methods under Scenario (S1) are given in Tables \ref{3changepointssparisty0.8}-\ref{50changepointssparisty0.8}. The method with the highest empirical frequency of $\hat{N} - N$ being equal to zero (or close to zero, depending on the example) and those within $5\%$ off the highest are given in bold. For Scenario (S2), the results are presented in Table \ref{allchangepointssparistytrend}.
\begin{table*}
\centering
\caption{Distribution of $\hat{N} - N$ over 100 simulated multivariate data sequences with 20 change-points under Scenario (S1) of Section \ref{subsec:pcm}. The average ARI, $d_H$, and computational times are also given}
\begin{tabular}{@{}llllllllll@{}}
 & & & \multicolumn{4}{|c|}{} & & & \\
Method & $sp$ & $d$ & \multicolumn{4}{|c|}{$\hat{N} - N$} & ARI & $d_h$ & Time (s)\\
 & & &$\leq -10$ & $(-10,-2)$ & $[-2,2]$ & $(2,10]$ & & & \\
$\boldsymbol{{\rm MID}_{\rm opt}}$& 0.2 & 30 & 0 & 4 & {\bf{96}} & 0 & 0.95 
& 0.19 & 0.72\\ 
DC & 0.2 & 30 & 2 & 98 & 0 & 0 &0.74 
&0.59 & 7.38\\ 
SBS & 0.2 & 30 & 100 & 0 & 0 & 0 & 0.50 
& 1.19& 1.36\\ 
INSPECT & 0.2 & 30 & 0 & 8 & 87 & 5 & 0.94 
& 0.23 & 0.44 \\ 
${\rm MIDPERL}_2$ & 0.2 & 30 & 0 & 16 & 84 & 0 & 0.94 & 0.25 & 476.36\\
$\boldsymbol{{\rm MIDPERL}_{\infty}}$ & 0.2 & 30 & 0 & 4 & {\bf{96}} & 0 & 0.95 & 0.17 & 110.60\\
\hline
$\boldsymbol{{\rm MID}_{\rm opt}}$& 0.2 & 100 & 0 & 1 & {\bf{98}} & 1 & 0.97 
& 0.08 & 2.20\\
DC & 0.2 & 100 & 16 & 84 & 0 & 0 & 0.76 
& 0.58 & 18.36\\ 
SBS & 0.2 & 100 & 100 & 0 & 0 & 0 & 0.57 
& 0.97 & 3.17\\ 
INSPECT & 0.2 & 100 & 0 & 3 & 91 & 6 & 0.97 
& 0.20 & 1.86 \\ 
$\boldsymbol{{\rm MIDPERL}_2}$ & 0.2 & 100 & 0 & 6 & {\bf{94}} & 0  & 0.97 & 0.21 & 1488.64\\ 
$\boldsymbol{{\rm MIDPERL}_{\infty}}$ & 0.2 & 100 & 0 & 1 & {\bf{99}} & 0 & 0.97 & 0.13 & 427.60\\
\hline
$\boldsymbol{{\rm MID}_{\rm opt}}$& 0.5 & 30 & 0 & 0 & {\bf{100}} & 0 & 0.97 
& 0.11 & 0.70 \\
DC & 0.5 & 30 & 17 & 83 & 0 & 0 & 0.75 
& 0.62 & 10.11\\ 
SBS & 0.5 & 30 & 100 & 0 & 0 & 0 & 0.45 
& 1.41 & 1.34\\ 
INSPECT & 0.5 & 30 & 0 & 1 & 94 & 5 & 0.97 
& 0.16 &0.44 \\ 
$\boldsymbol{{\rm MIDPERL}_2}$ & 0.5 & 30 & 0 & 1 & {\bf{99}} & 0 & 0.98 & 0.14 & 461.75\\
$\boldsymbol{{\rm MIDPERL}_{\infty}}$ & 0.5 & 30 & 0 & 0 & {\bf{99}} & 1 & 0.97 & 0.12 & 103.31\\
\hline
$\boldsymbol{{\rm MID}_{\rm opt}}$& 0.5 & 100 & 0 & 0 & {\bf{100}} & 0 & 0.98 
& 0.06 & 2.15
\\
DC & 0.5 & 100 & 22 & 78 & 0 & 0 & 0.76 
& 0.58 & 21.03\\ 
SBS & 0.5 & 100 & 100 & 0 & 0 & 0 & 0.49 
& 1.20 & 3.21\\ 
\textbf{INSPECT} & 0.5 & 100 & 0 & 0 & \textbf{95} & 5 & 0.98 
&0.11 & 0.99\\ 
$\boldsymbol{{\rm MIDPERL}_2}$ & 0.5 & 100 & 0 & 0 & {\bf{100}} & 0 & 0.98 & 0.12 &1471.95\\
$\boldsymbol{{\rm MIDPERL}_{\infty}}$ & 0.5 & 100 & 0 & 0 & {\bf{100}} & 0 & 0.97 & 0.09 &450.34\\
\hline
$\boldsymbol{{\rm MID}_{\rm opt}}$ & 0.8 & 30 & 0 & 2 & {\bf{98}} & 0 & 0.98 
&0.16 & 0.72\\ 
DC & 0.8 & 30 & 24 & 76 & 0 & 0 & 0.74 
&0.62 &11.74 \\ 
SBS & 0.8 & 30 & 100 & 0 & 0 & 0 & 0.41 
&1.53&1.34 \\ 
\textbf{INSPECT} & 0.8 & 30 & 0 & 0 & \textbf{97} & 3 & 0.98 
&0.10&0.44 \\ 
$\boldsymbol{{\rm MIDPERL}_2}$ & 0.8 & 30 & 0 & 1 & {\bf{99}} & 0 & 0.98& 0.14 &417.58\\
$\boldsymbol{{\rm MIDPERL}_{\infty}}$ & 0.8 & 30 & 0 & 0 & {\bf{100}} & 0 & 0.97 & 0.12& 95.89 \\
\hline
$\boldsymbol{{\rm MID}_{\rm opt}}$ & 0.8 & 100 & 0 &  2 & {\bf{98}} & 0 & 0.98 
& 0.13 & 3.66\\ 
DC & 0.8 & 100 & 16 & 84 & 0 & 0 & 0.77 
&0.57&21.20\\ 
SBS & 0.8 & 100 & 100 & 0 & 0 & 0 & 0.46 
&1.29&3.18\\ 
{\bf {INSPECT}} & 0.8 & 100 & 0 & 0 & {\bf{96}} & 4 & 0.99 
&0.08&1.86\\ 
$\boldsymbol{{\rm MIDPERL}_2}$ & 0.8 & 100 & 0 & 0 & {\bf{100}} & 0 & 0.99 & 0.09 & 1467.22\\
$\boldsymbol{{\rm MIDPERL}_{\infty}}$ & 0.8 & 100 & 0 & 0 & {\bf{100}} & 0 &0.98 & 0.07 &459.46\\
\end{tabular}
\label{20changepointssparisty0.8}
\end{table*}
\begin{table*}
\centering
\caption{Distribution of $\hat{N} - N$ over 100 simulated multivariate data sequences with 50 change-points under Scenario (S1) of Section \ref{subsec:pcm}. The average ARI, $d_H$, and computational times are also given}
\begin{tabular}{@{}llllllllll@{}}
 & & & \multicolumn{4}{|c|}{} & & & \\
Method & $sp$ & $d$ & \multicolumn{4}{|c|}{$\hat{N} - N$} & ARI & $d_h$ & Time (s) \\
 & & &$\leq -40$ & $(-40,-20)$ & $[-20,-10)$ & $[-10,10]$ & & & \\
$\boldsymbol{{\rm MID}_{\rm opt}}$& 0.2 & 30 & 0 & 0 & 1 & \textbf{99} & 0.88
& 0.52& 1.48\\ 
DC & 0.2 & 30 & 1 & 94 & 5 & 0 &0.56 
&1.29 & 10.41\\ 
SBS & 0.2 & 30 & 100 & 0 & 0 & 0 & 0.19 
& 4.34& 1.91\\ 
\textbf{INSPECT} & 0.2 & 30 & 0 & 0 & 0 & \textbf{100} &0.88
& 0.53& 0.76 \\ 
$\boldsymbol{{\rm MIDPERL}_2}$ & 0.2 & 30 & 0 & 0 &3 & {\bf{97}} & 0.88 & 0.50 & 650.05 \\
$\boldsymbol{{\rm MIDPERL}_{\infty}}$ & 0.2 & 30 & 0 & 0 & 3 & {\bf{97}} & 0.88 & 0.50 &101.91\\
\hline
$\boldsymbol{{\rm MID}_{\rm opt}}$& 0.2 & 100 & 0 & 0 & 0 & \textbf{100} & 0.93
&0.40 & 4.49\\
DC & 0.2 & 100 & 0 & 72 & 28 & 0 & 0.59 
& 1.13& 22.60\\ 
SBS & 0.2 & 100 & 98& 2& 0 & 0 & 0.23 
& 3.65 &4.62\\ 
\textbf{INSPECT} & 0.2 & 100 & 0 & 0 & 0 & \textbf{100} & 0.93 &0.51 & 2.87\\ 
$\boldsymbol{{\rm MIDPERL}_2}$ & 0.2 & 100 & 0 & 0 &1 & {\bf{99}} & 0.92 & 0.54 & 2188.56\\ 
$\boldsymbol{{\rm MIDPERL}_{\infty}}$ & 0.2 & 100 & 0 & 0 & 0 & {\bf{100}} & 0.91& 0.46 & 530.94\\
\hline 
$\boldsymbol{{\rm MID}_{\rm opt}}$& 0.5 & 30 & 0 & 0 & 0 & {\bf{100}} & 0.92 
&0.44&1.44 \\
DC & 0.5 & 30 & 1 & 79 & 20 & 0 &0.56 
&1.26&11.62\\ 
SBS & 0.5 & 30 & 100 & 0 & 0 & 0 & 0.16 
&5.02&1.85\\ 
{\bf{INSPECT}} & 0.5 & 30 & 0 & 0 & 0 & {\bf{100}} & 0.94 
&0.43&0.76 \\ 
$\boldsymbol{{\rm MIDPERL}_2}$ & 0.5 & 30 & 0 & 0  &0 & {\bf{100}} & 0.94 & 0.48 & 576.55\\
$\boldsymbol{{\rm MIDPERL}_{\infty}}$ & 0.5 & 30 & 0 & 0 & 0& {\bf{100}} & 0.92 & 0.43 & 85.13\\
\hline
$\boldsymbol{{\rm MID}_{\rm opt}}$& 0.5 & 100 & 0 & 0 & 0 & {\bf{100}} & 0.95 
&0.27&4.38
\\
DC & 0.5 & 100 & 0 & 87 & 13 & 0 & 0.59 
&1.07&28.03\\ 
SBS & 0.5 & 100 & 100 & 0 & 0 & 0 &0.18
&4.49&4.46\\ 
\textbf{INSPECT} & 0.5 & 100 & 0 & 0 & 0 & \textbf{100} & 0.97 
&0.36&2.86\\ 
$\boldsymbol{{\rm MIDPERL}_2}$ & 0.5 & 100 & 0 & 0&0 & {\bf{100}} & 0.96 & 0.42 & 2147.27 \\
$\boldsymbol{{\rm MIDPERL}_{\infty}}$ & 0.5 & 100 & 0 & 0 & 0& {\bf{100}} & 0.93 & 0.39 & 524.08\\
\hline
$\boldsymbol{{\rm MID}_{\rm opt}}$ & 0.8 & 30 & 0 & 0 & 0 & {\bf{100}} & 0.95
&0.48&2.13\\ 
DC & 0.8 & 30 & 0 & 83 & 17 & 0 & 0.58 
&1.12&13.46\\ 
SBS & 0.8 & 30 & 100 & 0 & 0 & 0 & 0.15 
&5.44&1.85 \\ 
\textbf{INSPECT} & 0.8 & 30 & 0 & 0 & 0 & \textbf{100} & 0.96 
&0.37&0.76\\ 
$\boldsymbol{{\rm MIDPERL}_2}$ & 0.8 & 100 & 0 & 0 &0 & {\bf{100}} & 0.96 & 0.42 & 513.94 \\
$\boldsymbol{{\rm MIDPERL}_{\infty}}$ & 0.8 & 100 & 0 & 0 & 0&  {\bf{100}} & 0.93 & 0.41 & 77.26\\
\hline
$\boldsymbol{{\rm MID}_{\rm opt}}$ & 0.8 & 100 & 0 &  0 & 0 & {\bf{100}} & 0.96 
&0.44&6.92\\ 
DC & 0.8 & 100 & 0 & 90 & 10 & 0 & 0.60 
&1.07&29.73\\ 
SBS & 0.8 & 100 & 100 & 0 & 0 & 0 & 0.17 
&5.02&4.41\\ 
{\bf {INSPECT}} & 0.8 & 100 & 0 & 0 & 0 & {\bf{100}} & 0.98
&0.29&2.86\\ 
$\boldsymbol{{\rm MIDPERL}_2}$ & 0.8 & 100 & 0 & 0 &0 & {\bf{100}} & 0.98 & 0.36 & 1723.12 \\
$\boldsymbol{{\rm MIDPERL}_{\infty}}$ & 0.8 & 100 & 0 & 0 & 0&  {\bf{100}} & 0.94 & 0.37 & 511.36\\
\end{tabular}
\label{50changepointssparisty0.8}
\end{table*}
As the tables show, MID performs extremely well in all setups in both (S1) and (S2). More specifically, for (S1) our method is 17 out of 18 times either the best method overall or within $5\%$ off the best method with respect to the estimated number of change-points. In all cases, MID attains very high values for the ARI, and quite small (in most cases it actually attains the smallest value) ones for the scaled Hausdorff distance; such results justify that apart from being extremely accurate in estimating the correct number of change-points, MID is also very accurate regarding the estimated change-point locations. Regarding the permutation-based variants of MID, both ${\rm MIDPERL}_2$ and ${\rm MIDPERL}_\infty$ show a very good behaviour in all different scenarios in terms of accuracy with respect to both the estimated number and the estimated change-point locations. Although these permutation-based variants do not require specification of the threshold, their computational cost is quite large (see Tables \ref{3changepointssparisty0.8}-\ref{50changepointssparisty0.8}).

In regards to the competing methods, INSPECT has a very accurate behaviour in all different scenarios. DC's performance is very good only in those cases with three change-points and it seems to heavily underestimate in cases with more, regularly occurring change-points. SBS does not have a good performance in the scenarios tested; it underestimates the number of change-points.

With respect to Scenario (S2) as in Section \ref{subsec:cplm}, the results in Table \ref{allchangepointssparistytrend} exhibit MID's very good performance. In conclusion, taking into account its low computational time, we can deduce that our proposed method is reliable and quick in accurately detecting change-points under various different (with respect to the number of change-points, their magnitude, the sparsity, the dimensionality of the data, and the structure of the changes) multivariate settings. \textsf{R} code and instructions to replicate the results in this section are available on \url{https://github.com/apapan08/Simulations-MID}.
\begin{table*}
\centering
\caption{Distribution of $\hat{N} - N$ over 100 simulated multivariate data sequences under Scenario (S2) of Section \ref{subsec:cplm}. We present the results of MID for different levels of sparsity and dimension for the data sequence. The number of change-point is equal to 3, 20, or 50. The average ARI, $d_H$, and computational times are also given}
\begin{tabular}{@{}llllllllllll@{}}
 & & & \multicolumn{6}{|c|}{} & & & \\
$\hat{N}$ & $sp$ & $d$ & \multicolumn{6}{|c|}{$\hat{N} - N$} & ARI & $d_h$ & Time (s)\\
 & & &$-2$ & $-1$ & 0 & 1 & 2 & 3 & & & \\
3& 0.2 & 10 & 0 & 0 & 92 & 7  & 1 & 0& 0.987 & 0.011& 0.293\\ 
3 & 0.5 & 10 & 0 & 0 & 92 & 6 & 2 &0&0.984 & 0.034 &0.251\\ 
3 & 0.8 & 10 & 0 & 0 &94 & 6 & 0 & 0 & 0.991 & 0.008& 0.223\\ 
\hline
3& 0.2 & 30 & 0 & 0 & 96 & 4 & 0 & 0 & 0.990& 0.010& 0.668\\ 
3 & 0.5 & 30 & 0 & 0 & 96 & 4 & 0 &0 & 0.987 &0.013 & 0.688\\ 
3 & 0.8 & 30 & 0 & 0 & 99 & 1 & 0 & 0 & 0.997 & 0.003& 0.647\\ 
\hline
3& 0.2 & 100 & 0 & 0 & 96 & 3 & 1 & 0&0.989& 0.015& 2.139\\ 
3 & 0.5 & 100 & 0 & 0 & 97 & 3 & 0 & 0 & 0.989 &0.008 & 2.091\\ 
3 & 0.8 & 100 & 0 & 0 & 99 & 1 & 0 & 0 & 0.999 & 0.001& 2.061\\ 
\hline
20& 0.2 & 10 & 0 & 0 & 87 & 11 & 2 & 0& 0.957& 0.045& 0.367\\ 
20 & 0.5 & 10 & 0 & 0 & 85 & 14 & 1 &0 & 0.960 &0.044 & 0.450\\ 
20 & 0.8 & 10 & 0 & 0 & 98 & 1 & 1 & 0 & 0.976 & 0.028& 0.459\\ 
\hline
20& 0.2 & 30 & 0 & 0 & 92 & 7 & 1 & 0& 0.962& 0.041& 1.080\\ 
20 & 0.5 & 30 & 0 & 0 & 82 & 16 & 2 &0 & 0.963 &0.046 &1.048\\ 
20 & 0.8 & 30 & 0 & 0 & 94 & 6 & 0 & 0 & 0.985 & 0.021& 0.983\\ 
\hline
20& 0.2 & 100 & 0 & 0 & 90 & 9 & 1 & 0& 0.964& 0.036& 3.775\\ 
20 & 0.5 & 100 & 0 & 0 & 85 & 14 & 1 &0 & 0.964 &0.036 & 3.495\\ 
20 & 0.8 & 100 & 0 & 3 & 97 & 0 & 0 & 0 & 0.994 & 0.189& 3.335\\ 
\hline
50& 0.2 & 10 & 0 & 3 & 86 &10 & 1 & 0 & 0.915& 0.136& 0.860\\ 
50 & 0.5 &10 &0 & 1 & 79 & 17 & 3 &0  & 0.921 &0.123 & 0.748\\ 
50 & 0.8 & 10 & 0 & 7 & 86 & 7 & 0 & 0 & 0.951 & 0.106& 0.712\\ 
\hline
50& 0.2 & 30 & 0 & 0 & 96 & 3 & 1 & 0 & 0.925& 0.101&2.165\\ 
50 & 0.5 & 30 & 0 & 0 & 84 & 14 & 1 & 1 &0.928 & 0.103 & 2.118\\ 
50 & 0.8 & 30 & 0 & 11 & 88 & 1 & 0 & 0 &0.970 & 0.094 & 2.035\\ 
\hline
50& 0.2 & 100 & 0 & 0 & 85 & 15 & 0 & 0 & 0.931& 0.095&6.879\\ 
50 & 0.5 & 100 & 0 & 0 & 82 & 17 & 1 & 0 & 0.934 &0.090 & 6.716\\ 
50 & 0.8 & 100 & 2 & 12 & 86 & 0 & 0 & 0 & 0.987 & 0.086& 6.784\\ 
\end{tabular}
\label{allchangepointssparistytrend}
\end{table*}

\section{Real data examples}
\label{sec:Real_data}
\vspace{-0.05in}
\subsection{UK House Price Index}
\label{subsec:hpi}
\begin{figure*}
  \makebox[\textwidth][c]{\includegraphics[width=0.9\textwidth, height = 0.35\textheight]{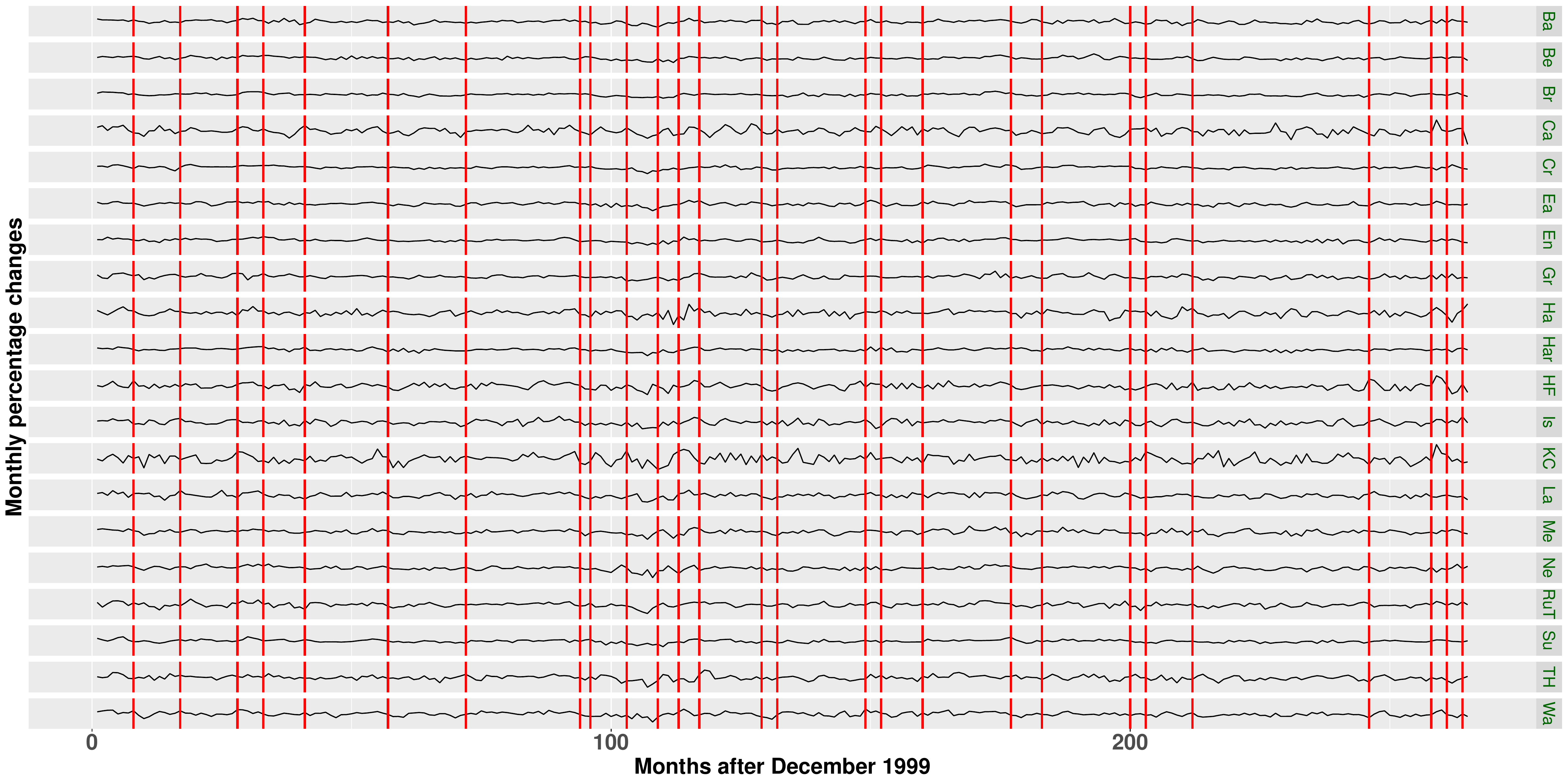}}%
  \caption{The monthly percentage changes in the UK house price index for the twenty London boroughs under consideration. The estimated change-point locations when ${\rm MID}_{{\rm opt}}$ was employed can be seen with red, vertical lines.}
  \label{fig:HPI_data_multi}
\end{figure*}

In this section, the performance of our method is studied on monthly percentage changes in the UK house price index for all property types from January 2000 to January 2022 in twenty London Boroughs. The data are available from \url{http://landregistry.data.gov.uk/app/ukhpi} and they were last accessed in May 2022. We have a multivariate data sequence $\boldsymbol{X_t} = \left(X_{t,1}, \ldots, X_{t,20}\right)^{\intercal}$, with $t = 1,\ldots, 265$. The boroughs used are : Barnet (Ba), Bexley (Be), Bromley (Br), Camden (Ca), Croydon (Cr), Ealing (Ea), Enfield (En), Greenwich (Gr), Hackney (Ha), Hammersmith and Fulham (HF), Harrow (Har), Islington (Is), Kensington and Chelsea (KC), Lambeth (La), Merton (Me), Newham (Ne), Richmond upon Thames (RuT), Sutton (Su), Tower Hamlets (TH), and Wandsworth (Wa). Similar data have been investigated in \cite{NOT}, \cite{fryzlewicz2014wild}, \cite{fryzlewicz2020wild}, and \cite{anastasiou2019detecting}; however under a univariate setting. Figure \ref{fig:HPI_data_multi} indicates the results when ${\rm MID}_{{\rm opt}}$ of Section \ref{subsec:reduction} was employed for the detection of changes under Scenario (S1) as explained in Section \ref{subsec:theory}; with respect to the threshold constant, we used the optimal values as in Table \ref{tab:best_thres} for $\alpha = 0.05$.
Our method identifies 27 change-points in the mean structure of the multivariate data sequence at hand. We have analysed the same data using the competing methods of Section \ref{sec:simulations}. INSPECT detects 26 change-points at locations very similar to those detected by our method, DC detects one change-point at location 41, while SBS does not detect any change-points. We need to highlight, though, that in INSPECT, multiple change-points are estimated using a wild binary segmentation scheme, which, due to the randomness involved, does not necessarily detect the same change-points when it is employed more than once over the same data set.
\subsection{The COVID-19 outbreak}
\label{subsec:covid}
The performance of our method is also investigated on data from the COVID-19 pandemic. In this case, we focus though on the detection of changes under Scenario (S2) as described in Section \ref{subsec:theory}. The data under consideration consist of the daily number of new lab-confirmed COVID-19 cases in the four constituent countries of United Kingdom; England, Northern Ireland, Scotland, and Wales. The period under investigation is from 01/04/2020 until 01/04/2022. The data are available from \url{https://coronavirus.data.gov.uk} and they were last accessed on the $30^{{\rm th}}$ of April 2022. Based on the description, in this example we have a multivariate data sequence $\boldsymbol{X_t} = \left(X_{t,1}, \ldots, X_{t,4}\right)^{\intercal}$, with $t = 1,\ldots, 731$. Due to the fact that the data are positive integer numbers, we perform the Anscombe transform, $a:\mathbb{N} \rightarrow \mathbb{R}$, with $a(x) = 2\sqrt{x+3/8}$, to each $X_{t,j}$; this transform brings the distribution of the component data sequences closer to the Gaussian one with constant variance.

Our method has detected 29 change-points in the vector of the first partial derivatives (changes in the slope for the component univariate data sequences) which seem to capture the important movements in the data.
\begin{figure*}
  \makebox[\textwidth][c]{\includegraphics[width=0.9\textwidth, height = 0.35\textheight]{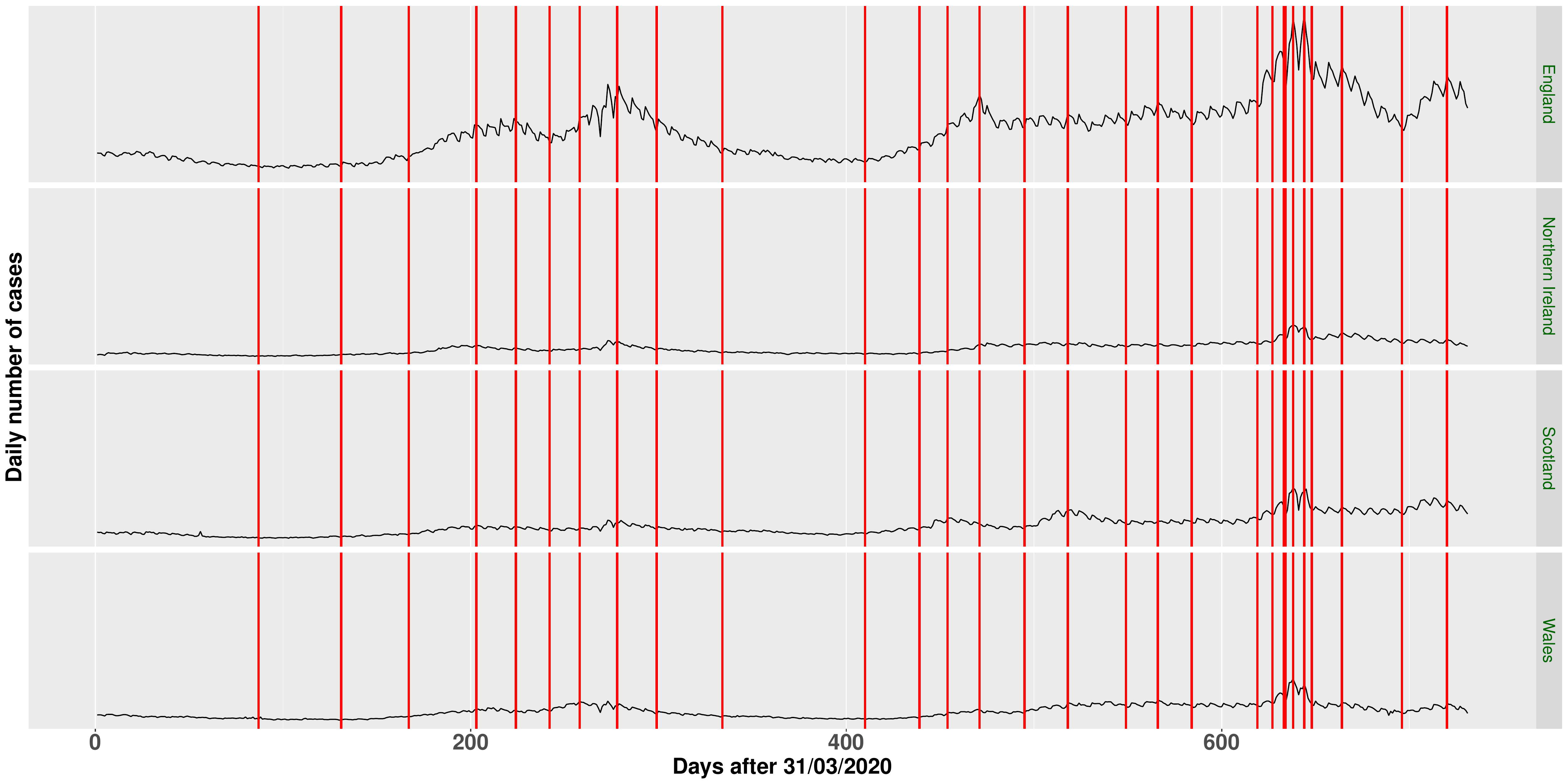}}%
  \caption{The daily number of COVID-19 cases for England, Northern Ireland, Scotland, and Wales. The estimated change-point locations in the trend of each component data sequence when our proposed ${\rm MID}_{{\rm opt}}$ method was employed can be seen with red, vertical lines.}
  \label{fig:covid_data_multi}
\end{figure*}
\section{Conclusion}
\label{sec:conclusion}
In this paper, the MID methodology has been proposed for multiple generalized change-point detection in multivariate, possibly high-dimensional, data sequences. The algorithm partly constitutes a generalization of the Isolate-Detect algorithm of \cite{anastasiou2019detecting}, which has been developed for univariate change-point detection. Mean-dominant norms, as those in \eqref{mean_dominant}, are employed for the aggregation of the information across the different components of the multivariate data sequence.

The aggregated values for the relevant contrast function (depending on the structure of the changes) are collected and compared to a threshold value, $\zeta_{T,d}$. The rate of $\zeta_{T,d}$ with respect to both the length, $T$, of the data sequence and its dimensionality, $d$, has been proven to be $\mathcal{O}\left(\sqrt{\log(Td^{1/4})}\right)$. The optimal multiplicative threshold constant, $C$, so that $\zeta_{T,d} = C\sqrt{\log(Td^{1/4})}$ has been derived through a large scale simulation study, with special attention in controlling the Type I error rate, $\alpha$, of falsely detecting change-points for various values of the dimensionality of the data sequence.

Misspecification of the threshold is possible and could lead to misestimation of the underlying signal. To solve such issues, in Section \ref{subsec:permutation}, a permutation-based variant of our MID algorithm has been proposed with no threshold choice requirement. The algorithmic steps of MID remain the same; the difference lies in the way the method chooses to accept or reject a change-point within the interval under consideration, suppose this is denoted by $I=[s^*,e^*]$, where $1 \leq s^* < e^* \leq T$. In Section \ref{methodology}, it has been explained in detail that MID will first return a vector $\boldsymbol{v} \in \mathbb{R}^{J}$, where $J$ is the amount of all change-point candidates. The elements of $\boldsymbol{v}$ correspond to the aggregated contrast function values for each candidate point in $I$. The next step is to store the maximum value of $\boldsymbol{v}$ as obtained from the original data and compare it to the appropriate quantiles of the empirical distribution of the values obtained when applying the same steps to several permuted versions of the data; see the exact steps in Section  \ref{subsec:permutation}. The proposed permutation-based variant of MID, though computationally expensive, performs very well in terms of accuracy with respect to the estimated number and locations of the change-points; see the results in Section \ref{sec:simulations}.

The choice of the mean-dominant norm to be employed in the aggregation step of MID has already been discussed in the current paper, more specifically in Section \ref{subsec:reduction}. Our aim has been to provide a method that in practice would require minimal parameter choice to the user; towards this purpose, a data-adaptive variant, named ${\rm MID}_{{\rm opt}}$, of the method has been constructed. We first estimate the sparsity level in the given multivariate data; the steps are explained in detail in Section \ref{subsec:reduction}. Depending on the value of the estimated sparsity, ${\rm MID}_{\rm opt}$ estimates the change-points employing either the $L_{\infty}$ or the $L_2$ mean-dominant norm as defined in \eqref{mean_dominant}.

Through simulated and real-life data examples presented in Sections \ref{sec:simulations} and \ref{sec:Real_data}, respectively, it has been shown that MID has very good performance in terms of accuracy and speed. Specifically, MID lies in the top 5\% (in terms of the accurate estimation of the number and the location of the change-points) of the best methods when compared to the state-of-the-art competitors. In addition, ${\rm MID}_{\rm opt}$ is a very quick detection method which in a few seconds can analyse signals with length in the range of thousands and dimensionality in the range of hundreds; this is carried out automatically with minimal decision making from the user on the aggregation method to be employed.
\vspace{0.1in}
\\
\raggedright{{\textbf{Acknowledgments}}}\\
This research occurred whilst Angelos Papanastasiou was employed as a Research Special Scientist at the University of Cyprus, supported by the start-up funding that Andreas Anastasiou has received from the University of Cyprus.
\bibliographystyle{Chicago}
\bibliography{General_version}

\begin{thebibliography}{}

\bibitem[\protect\citeauthoryear{Anastasiou, Cribben, and
  Fryzlewicz}{Anastasiou et~al.}{2022}]{CCID}
Anastasiou, A., I.~Cribben, and P.~Fryzlewicz (2022).
\newblock Cross-covariance isolate detect: {A} new change-point method for
  estimating dynamic functional connectivity.
\newblock {\em Medical Image Analysis\/}~{\em 75}.

\bibitem[\protect\citeauthoryear{Anastasiou and Fryzlewicz}{Anastasiou and
  Fryzlewicz}{2022}]{anastasiou2019detecting}
Anastasiou, A. and P.~Fryzlewicz (2022).
\newblock Detecting multiple generalized change-points by isolating single
  ones.
\newblock {\em Metrika\/}~{\em 85}, 141--174.

\bibitem[\protect\citeauthoryear{Antoch and Hu{\v{s}}kov{\'a}}{Antoch and
  Hu{\v{s}}kov{\'a}}{2001}]{antoch2001permutation}
Antoch, J. and M.~Hu{\v{s}}kov{\'a} (2001).
\newblock Permutation tests in change point analysis.
\newblock {\em Statistics \& probability letters\/}~{\em 53\/}(1), 37--46.

\bibitem[\protect\citeauthoryear{Auger and Lawrence}{Auger and
  Lawrence}{1989}]{auger_lawrence}
Auger, I.~E. and C.~E. Lawrence (1989).
\newblock Algorithms for the {O}ptimal {I}dentification of {S}egment
  {N}eighborhoods.
\newblock {\em Bulletin of Mathematical Biology\/}~{\em 51}, 39--54.

\bibitem[\protect\citeauthoryear{Baranowski, Chen, and Fryzlewicz}{Baranowski
  et~al.}{2019}]{NOT}
Baranowski, R., Y.~Chen, and P.~Fryzlewicz (2019).
\newblock Narrowest-over-threshold detection of multiple change points and
  change-point-like features.
\newblock {\em Journal of the Royal Statistical Society: Series B\/}~{\em 81},
  649--672.

\bibitem[\protect\citeauthoryear{B{\"u}cher, Kojadinovic, Rohmer, and
  Segers}{B{\"u}cher et~al.}{2014}]{bucher2014detecting}
B{\"u}cher, A., I.~Kojadinovic, T.~Rohmer, and J.~Segers (2014).
\newblock Detecting changes in cross-sectional dependence in multivariate time
  series.
\newblock {\em Journal of Multivariate Analysis\/}~{\em 132}, 111--128.

\bibitem[\protect\citeauthoryear{Cabrieto, Tuerlinckx, Kuppens, Hunyadi, and
  Ceulemans}{Cabrieto et~al.}{2018}]{cabrieto2018testing}
Cabrieto, J., F.~Tuerlinckx, P.~Kuppens, B.~Hunyadi, and E.~Ceulemans (2018).
\newblock Testing for the presence of correlation changes in a multivariate
  time series: A permutation based approach.
\newblock {\em Scientific reports\/}~{\em 8\/}(1), 1--20.

\bibitem[\protect\citeauthoryear{Carlstein}{Carlstein}{1988}]{Carlstein}
Carlstein, E. (1988).
\newblock Nonparametric change-point estimation.
\newblock {\em Annals of Statistics\/}~{\em 16}, 188--197.

\bibitem[\protect\citeauthoryear{Cho}{Cho}{2016}]{cho2016change}
Cho, H. (2016).
\newblock Change-point detection in panel data via double cusum statistic.
\newblock {\em Electronic Journal of Statistics\/}~{\em 10\/}(2), 2000--2038.

\bibitem[\protect\citeauthoryear{Cho and Fryzlewicz}{Cho and
  Fryzlewicz}{2015}]{cho2015multiple}
Cho, H. and P.~Fryzlewicz (2015).
\newblock Multiple-change-point detection for high dimensional time series via
  sparsified binary segmentation.
\newblock {\em Journal of the Royal Statistical Society: Series B: Statistical
  Methodology\/}, 475--507.

\bibitem[\protect\citeauthoryear{Cho and Kirch}{Cho and
  Kirch}{2020}]{Cho_Kirch_2020}
Cho, H. and C.~Kirch (2020).
\newblock Data segmentation algorithms: {U}nivariate mean change and beyond.
\newblock {\em arXiv preprint arXiv:2012.12814\/}.

\bibitem[\protect\citeauthoryear{Enikeeva and Harchaoui}{Enikeeva and
  Harchaoui}{2019}]{enikeeva2013high}
Enikeeva, F. and Z.~Harchaoui (2019).
\newblock High-dimensional change-point detection under sparse alternatives.
\newblock {\em Annals of Statistics\/}~{\em 47\/}(4), 2051--2079.

\bibitem[\protect\citeauthoryear{Fearnhead, Maidstone, and Letchford}{Fearnhead
  et~al.}{2019}]{CPOP}
Fearnhead, P., R.~Maidstone, and A.~Letchford (2019).
\newblock Detecting {C}hanges in {S}lope {W}ith an ${L}_0$ {P}enalty.
\newblock {\em Journal of Computational and Graphical Statistics\/}~{\em {\bf
  28}}, 265--275.

\bibitem[\protect\citeauthoryear{Frick, Munk, and Sieling}{Frick
  et~al.}{2014}]{frick2014multiscale}
Frick, K., A.~Munk, and H.~Sieling (2014).
\newblock Multiscale change point inference.
\newblock {\em Journal of the Royal Statistical Society: Series B: Statistical
  Methodology\/}, 495--580.

\bibitem[\protect\citeauthoryear{Fryzlewicz}{Fryzlewicz}{2014}]{fryzlewicz2014wild}
Fryzlewicz, P. (2014).
\newblock Wild binary segmentation for multiple change-point detection.
\newblock {\em Annals of Statistics\/}~{\em 42}, 2243--2281.

\bibitem[\protect\citeauthoryear{Fryzlewicz}{Fryzlewicz}{2020}]{fryzlewicz2020wild}
Fryzlewicz, P. (2020).
\newblock Detecting possibly frequent change-points: Wild binary segmentation 2
  and steepest-drop model selection.
\newblock {\em Journal of the Korean Statistical Society\/}~{\em 49},
  1027--1070.

\bibitem[\protect\citeauthoryear{Groen, Kapetanios, and Price}{Groen
  et~al.}{2013}]{groen2013multivariate}
Groen, J.~J., G.~Kapetanios, and S.~Price (2013).
\newblock Multivariate methods for monitoring structural change.
\newblock {\em Journal of Applied Econometrics\/}~{\em 28\/}(2), 250--274.

\bibitem[\protect\citeauthoryear{Hampel}{Hampel}{1974}]{Hampel}
Hampel, F.~R. (1974).
\newblock The influence curve and its role in robust estimation.
\newblock {\em Journal of the American Statistical Association\/}~{\em 69},
  383--393.

\bibitem[\protect\citeauthoryear{Hocking, Schleiermacher, Janoueix-Lerosey,
  Boeva, Cappo, Delattre, Bach, and Vert}{Hocking
  et~al.}{2013}]{hocking2013learning}
Hocking, T.~D., G.~Schleiermacher, I.~Janoueix-Lerosey, V.~Boeva, J.~Cappo,
  O.~Delattre, F.~Bach, and J.-P. Vert (2013).
\newblock Learning smoothing models of copy number profiles using breakpoint
  annotations.
\newblock {\em BMC bioinformatics\/}~{\em 14\/}(1), 1--15.

\bibitem[\protect\citeauthoryear{Horv{\'a}th and Hu{\v{s}}kov{\'a}}{Horv{\'a}th
  and Hu{\v{s}}kov{\'a}}{2012}]{horvath2012change}
Horv{\'a}th, L. and M.~Hu{\v{s}}kov{\'a} (2012).
\newblock Change-point detection in panel data.
\newblock {\em Journal of Time Series Analysis\/}~{\em 33\/}(4), 631--648.

\bibitem[\protect\citeauthoryear{Hubert and Arabie}{Hubert and
  Arabie}{1985}]{hubert1985comparing}
Hubert, L. and P.~Arabie (1985).
\newblock Comparing partitions.
\newblock {\em Journal of classification\/}~{\em 2\/}(1), 193--218.

\bibitem[\protect\citeauthoryear{Incl\'an and Tiao}{Incl\'an and
  Tiao}{1994}]{Inclan_Tiao}
Incl\'an, C. and G.~C. Tiao (1994).
\newblock Use of cumulative sums of squares for retrospective detection of
  changes of variance.
\newblock {\em Journal of the American Statistical Association\/}~{\em 89},
  913--923.

\bibitem[\protect\citeauthoryear{Jackson, Sargle, Barnes, Arabhi, Alt,
  Gioumousis, Gwin, Sangtrakulcharoen, Tan, and Tsai}{Jackson
  et~al.}{2005}]{jackson}
Jackson, B., J.~D. Sargle, D.~Barnes, S.~Arabhi, A.~Alt, P.~Gioumousis,
  E.~Gwin, P.~Sangtrakulcharoen, L.~Tan, and T.~T. Tsai (2005).
\newblock An {A}lgorithm for {O}ptimal {P}artitioning of {D}ata on an
  {I}nterval.
\newblock {\em IEEE Signal Processing Letters\/}~{\em 12}, 105--108.

\bibitem[\protect\citeauthoryear{Jirak}{Jirak}{2015}]{jirak2015uniform}
Jirak, M. (2015).
\newblock Uniform change point tests in high dimension.
\newblock {\em Annals of Statistics\/}~{\em 43\/}(6), 2451--2483.

\bibitem[\protect\citeauthoryear{Killick, Fearnhead, and Eckley}{Killick
  et~al.}{2012}]{killick2012optimal}
Killick, R., P.~Fearnhead, and I.~A. Eckley (2012).
\newblock Optimal detection of changepoints with a linear computational cost.
\newblock {\em Journal of the American Statistical Association\/}~{\em
  107\/}(500), 1590--1598.

\bibitem[\protect\citeauthoryear{Kov\'{a}cs, Li, B\"{u}hlmann, and
  Munk}{Kov\'{a}cs et~al.}{2020}]{seeded}
Kov\'{a}cs, S., H.~Li, P.~B\"{u}hlmann, and A.~Munk (2020).
\newblock Seeded {B}inary {S}egmentation: {A} general methodology for fast and
  optimal change point detection.
\newblock {\em arXiv preprint arXiv:2002.06633\/}.

\bibitem[\protect\citeauthoryear{Lavielle and Teyssiere}{Lavielle and
  Teyssiere}{2006}]{lavielle2006detection}
Lavielle, M. and G.~Teyssiere (2006).
\newblock Detection of multiple change-points in multivariate time series.
\newblock {\em Lithuanian Mathematical Journal\/}~{\em 46\/}(3), 287--306.

\bibitem[\protect\citeauthoryear{Lavielle and Teyssiere}{Lavielle and
  Teyssiere}{2007}]{lavielle2007adaptive}
Lavielle, M. and G.~Teyssiere (2007).
\newblock Adaptive detection of multiple change-points in asset price
  volatility.
\newblock In {\em Long memory in economics}, pp.\  129--156. Springer.

\bibitem[\protect\citeauthoryear{Maeng and Fryzlewicz}{Maeng and
  Fryzlewicz}{2019}]{Maeng_Fryzlewicz}
Maeng, H. and P.~Fryzlewicz (2019).
\newblock Detecting linear trend changes and point anomalies in data sequences.
\newblock {\em arXiv preprint arXiv:1906.01939, 2019\/}.

\bibitem[\protect\citeauthoryear{Matteson and James}{Matteson and
  James}{2014}]{matteson}
Matteson, D.~S. and N.~A. James (2014).
\newblock A {N}onparametric {A}pproach for {M}ultiple {C}hange {P}oint
  {A}nalysis of {M}ultivariate {D}ata.
\newblock {\em Journal of the American Statistical Association\/}~{\em {\bf
  109}}, 334--345.

\bibitem[\protect\citeauthoryear{Ombao, Von~Sachs, and Guo}{Ombao
  et~al.}{2005}]{ombao2005slex}
Ombao, H., R.~Von~Sachs, and W.~Guo (2005).
\newblock Slex analysis of multivariate nonstationary time series.
\newblock {\em Journal of the American Statistical Association\/}~{\em
  100\/}(470), 519--531.

\bibitem[\protect\citeauthoryear{Picard, Lebarbier, Hoebeke, Rigaill, Thiam,
  and Robin}{Picard et~al.}{2011}]{picard2011joint}
Picard, F., E.~Lebarbier, M.~Hoebeke, G.~Rigaill, B.~Thiam, and S.~Robin
  (2011).
\newblock Joint segmentation, calling, and normalization of multiple cgh
  profiles.
\newblock {\em Biostatistics\/}~{\em 12\/}(3), 413--428.

\bibitem[\protect\citeauthoryear{Rigaill}{Rigaill}{2015}]{rigaill}
Rigaill, G. (2015).
\newblock A pruned dynamic programming algorithm to recover the best
  segmentations with 1 to {K}max change-points.
\newblock {\em Journal de la Soci\'et\'e Fran\c caise de Statistique\/}~{\em
  156}, 180--205.

\bibitem[\protect\citeauthoryear{Schr{\"o}der and Fryzlewicz}{Schr{\"o}der and
  Fryzlewicz}{2013}]{schroder2013adaptive}
Schr{\"o}der, A.~L. and P.~Fryzlewicz (2013).
\newblock Adaptive trend estimation in financial time series via multiscale
  change-point-induced basis recovery.
\newblock {\em Statistics and its interface\/}~{\em 6\/}(4), 449--461.

\bibitem[\protect\citeauthoryear{Siris and Papagalou}{Siris and
  Papagalou}{2004}]{siris2004application}
Siris, V.~A. and F.~Papagalou (2004).
\newblock Application of anomaly detection algorithms for detecting syn
  flooding attacks.
\newblock In {\em IEEE Global Telecommunications Conference, 2004.
  GLOBECOM'04.}, Volume~4, pp.\  2050--2054. IEEE.

\bibitem[\protect\citeauthoryear{Tibshirani}{Tibshirani}{2014}]{Tibshirani}
Tibshirani, R.~J. (2014).
\newblock Adaptive piecewise polynomial estimation via trend filtering.
\newblock {\em Annals of Statistics\/}~{\em {\bf 42}}, 285--323.

\bibitem[\protect\citeauthoryear{Truong, Oudre, and Vayatis}{Truong
  et~al.}{2020}]{truong2020selective}
Truong, C., L.~Oudre, and N.~Vayatis (2020).
\newblock Selective review of offline change point detection methods.
\newblock {\em Signal Processing\/}~{\em 167}, 107299.

\bibitem[\protect\citeauthoryear{Vert and Bleakley}{Vert and
  Bleakley}{2010}]{vert2010fast}
Vert, J.-P. and K.~Bleakley (2010).
\newblock Fast detection of multiple change-points shared by many signals using
  group lars.
\newblock In {\em NIPS}, pp.\  2343--2351.

\bibitem[\protect\citeauthoryear{Wang and Samworth}{Wang and
  Samworth}{2018}]{wang2016high}
Wang, T. and R.~J. Samworth (2018).
\newblock High-dimensional changepoint estimation via sparse projection.
\newblock {\em Journal of the Royal Statistical Society: Series B\/}~{\em 80},
  57--83.

\bibitem[\protect\citeauthoryear{Yu}{Yu}{2020}]{Yu2020}
Yu, Y. (2020).
\newblock A review on minimax rates in change point detection and localisation.
\newblock {\em arXiv preprint arXiv:2011.01857\/}.

\bibitem[\protect\citeauthoryear{Yuan and Lin}{Yuan and
  Lin}{2006}]{yuan2006model}
Yuan, M. and Y.~Lin (2006).
\newblock Model selection and estimation in regression with grouped variables.
\newblock {\em Journal of the Royal Statistical Society: Series B (Statistical
  Methodology)\/}~{\em 68\/}(1), 49--67.

\bibitem[\protect\citeauthoryear{Zou, Yin, Feng, and Wang}{Zou
  et~al.}{2014}]{Zouetal2014}
Zou, C., G.~Yin, L.~Feng, and Z.~Wang (2014).
\newblock Nonparametric maximum likelihood approach to multiple change-point
  problems.
\newblock {\em The Annals of Statistics\/}~{\em {\bf 42}}, 970--1002.

\end{thebibliography}

\appendix
\section{Proof of Theorem \ref{theorem_consistency_S1}}
\label{sec:Proof_S1}
\vspace{-0.05in}
With $\tilde{X}_{s,e}^{b,j}$ as in \eqref{cusum_formula}, we denote, for $s \leq b < e$, by $\tilde{f}_{s,e}^{b,j} = \sqrt{\frac{e-b}{n(b-s+1)}}\sum_{t=s}^{b}f_{t,j} - \sqrt{\frac{b-s+1}{n(e-b)}}\sum_{t=b+1}^{e}f_{t,j}$ the value of the CUSUM statistic for the $j^{th}$ underlying component signal. Furthermore,
\begin{align}
\label{set_A_T}
\nonumber & C^{b}_{s,e}  = \max\limits_{1 \leq j \leq d} \left|\tilde{X}_{s,e}^{b,j}\right|, \qquad\quad D_{s,e}^{b} = \max\limits_{1 \leq j \leq d} \left|\tilde{f}_{s,e}^{b,j}\right|,\\
\nonumber & A_T^*  =\bigg\{\max_{1\leq s \leq b < e\leq T}\left\lbrace\max_{1 \leq j \leq d}\left|\tilde{X}_{s,e}^{t,j}-\tilde{f}_{s,e}^{t,j}\right|\right\rbrace\leq \sqrt{8\operatorname{log}\left(Td^{\frac{1}{4}}\right)}\bigg\},\\
& A_T  =\bigg\{\max_{1 \leq s \leq b < e \leq T}\left|C^{b}_{s,e}-D_{s,e}^{b}\right|\leq \sqrt{8\operatorname{log}\left(Td^{\frac{1}{4}}\right)}\bigg\}.
\end{align}
\begin{proof}
\textbf{Step 1:} We will show that $$\Prob\left(A_T\right) \geq 1 - 1/(12\sqrt{\pi}T).$$ Due to the fact that
\begin{align}
\nonumber \left|C^{b}_{s,e}-D_{s,e}^{b}\right| &=\bigg|\max_{1 \leq j \leq d}\left|\tilde{X}_{s,e}^{b,j}\right|-\max_{1 \leq j \leq d}\left|\tilde{f}_{s,e}^{b,j}\right|\bigg|\\
\nonumber & \leq \max_{1 \leq j \leq d}\left|\tilde{X}_{s,e}^{b,j}-\tilde{f}_{s,e}^{b,j}\right|,
\end{align}
it is straightforward to see that $\Prob\left(A_T\right) \geq \Prob\left(A_T^*\right)$ and therefore, it suffices to show that $\Prob\left((A^*_T)^{c}\right) \leq  1/(12\sqrt{\pi}T)$. From the definition of our model in \eqref{eq:model}, we have that for any $j \in \left\lbrace 1,\ldots, d\right\rbrace$, $\tilde{X}_{s,e}^{b,j} - \tilde{f}^{b,j}_{s,e} = \tilde{\epsilon}_{s,e}^{b,j}$, and simple calculations lead to $\tilde{\epsilon}_{s,e}^{t,j} \sim \mathcal{N}(0,\,1)$. Therefore, for $Z_j \sim \mathcal{N}(0,\,1)$, the Bonferroni inequality yields
\begin{align}
\label{proofA_T}
\nonumber    \Prob\left((A^*_T)^{c}\right) &=
\Prob\bigg( \max_{1 \leq s \leq b < e \leq T}\left\lbrace\max_{1 \leq j \leq d}|Z_{j}|\right\rbrace > \sqrt{8\operatorname{log}\left(Td^{\frac{1}{4}}\right)}\bigg) \\
\nonumber &\leq \sum_{1 \leq s \leq b < e\leq T}\Prob\bigg(\max_{1 \leq j \leq d}|Z_{j}|>\sqrt{8\operatorname{log}\left(Td^{\frac{1}{4}}\right)}\bigg)\\
\nonumber & \leq \frac{T^3}{6}\Prob\bigg(\max_{1 \leq j \leq d}|Z_{j}|>\sqrt{8\operatorname{log}\left(Td^{\frac{1}{4}}\right)}\bigg) \\
\nonumber &\leq \frac{T^3}{6}\sum_{j=1}^{d}\Prob\bigg(|Z_{j}|>\sqrt{8\operatorname{log}\left(Td^{\frac{1}{4}}\right)}\bigg)\\
\nonumber & = \frac{T^3}{3}\sum_{j=1}^{d}\Prob\bigg(Z_{j}>\sqrt{8\operatorname{log}\left(Td^{\frac{1}{4}}\right)}\bigg) \\
\nonumber &= \frac{dT^3}{3}\frac{\phi\bigg(\sqrt{8\operatorname{log}\left(Td^{\frac{1}{4}}\right)}\bigg)}{\sqrt{8\operatorname{log}\left(Td^{\frac{1}{4}}\right)}}=\frac{1}{3T\sqrt{2\pi}\sqrt{8\operatorname{log}(Td^{\frac{1}{4}})}}\\
& \leq \frac{1}{12\sqrt{\pi}T},
\end{align}
where $\phi(x), x \in \mathbb{R}$ is the probability density function of the standard normal distribution. We conclude that $\Prob(A_T^*) \geq 1 - 1/(12\sqrt{\pi}T)$, which leads to $\Prob(A_T) \geq 1 - 1/(12\sqrt{\pi}T)$.
\\
\textbf{Step 2 :} Let us denote by $\boldsymbol{\psi_{s,e}^b} = (\psi_{s,e}^b(1), \ldots, \psi_{s,e}^b(T))$ the contrast vector, where
\begin{equation}
\nonumber \psi_{s,e}^b(t) = \begin{cases}
    \sqrt{\frac{e-b}{n(b-s+1)}}, & t = s,\ldots,b,\\
    -\sqrt{\frac{b-s+1}{n(e-b)}}, & t=b+1,\ldots,e,\\
    0, & \text{otherwise},
  \end{cases},
\end{equation}
for $n = e-s+1$. With $\boldsymbol{f_j} = \left(f_{1,j}, \ldots, f_{T,j}\right)$ and for $[s,e]$ being any interval that contains only one true change-point, namely $r_j$, allow us for ease of presentation to introduce the notation
\begin{equation}
\label{A_seb}
    A_{s,e}^{b}(k,r_j) := \frac{\left\langle\boldsymbol{\psi_{s, e}^{b}}\left\langle \boldsymbol{f_{k}}, \boldsymbol{\psi_{s, e}^{b}}\right\rangle-\boldsymbol{\psi_{s,e}^{r_{j}}}\left\langle\boldsymbol{f_{k}},\boldsymbol{\psi_{s,e}^{r_{j}}}\right\rangle,\boldsymbol{\epsilon}\right\rangle}{\left\|\boldsymbol{\psi_{s, e}^{b}}\left\langle \boldsymbol{f_{k}}, \boldsymbol{\psi_{s, e}^{b}}\right\rangle-\boldsymbol{\psi_{s,e}^{r_{j}}}\left\langle \boldsymbol{f_{k}},\boldsymbol{\psi_{s,e}^{r_{j}}}\right\rangle\right\|_{2}}.
\end{equation}
In \eqref{A_seb} above, $k \in \left\lbrace 1, \ldots, d\right\rbrace$, while $\|\cdot\|_2$ is the Euclidean norm. Due to $\boldsymbol{\epsilon_t}, t = 1,\ldots, T$ following the $d$-variate standard normal distribution, then straightforward calculations lead to 
\begin{equation}
\label{distribution_A}
A_{s,e}^{b}(k,r_j) \sim \mathcal{N}(0,1).
\end{equation}
For
\begin{equation}
\label{set_B_T}
B_{T}=\left\{\max_{\substack{j=1, \ldots, N\\k=1,\ldots,d}}\max_{\substack{
r_{j-1}<s \leq r_{j}\\
r_{j} < e \leq r_{j+1} \\
s \leq b <e}}\left|A_{s,e}^{b}(k,r_j)\right| \leq \sqrt{8\operatorname{log}\left(Td^{\frac{1}{4}}\right)}\right\},
\end{equation} we will show that $\Prob\left(B_T\right) \geq 1 - 1/(12\sqrt{\pi}T)$. Using \eqref{distribution_A} and the Bonferroni inequality, then, for $Z \sim \mathcal{N}(0,1)$, similar steps as in \eqref{proofA_T} yield
\begin{equation}
\label{proofB_T}
\Prob\left(B_{T}^{c}\right) \leq \frac{1}{12\sqrt{\pi}T}.
\end{equation}
From \eqref{proofA_T} and \eqref{proofB_T} in Steps 1 and 2, respectively, we conclude that $\Prob\left(A_T \cap B_T\right) \geq 1 - \frac{1}{6\sqrt{\pi}T}.$\\
{\textbf{Step 3:}} This is the main part of our proof, where we explain in detail how to get the result in \eqref{mainresult_theorem}. For ease of understanding, we split this step into two smaller parts. From now on, we assume that $A_T^*$ (and therefore, also, $A_T$) and $B_T$ both hold. The constants we use are
\begin{equation}
\nonumber
C_1 = \sqrt{C_3} + \sqrt{8},\;\; C_2 = \frac{1}{\sqrt{6}} - \frac{2\sqrt{2}}{\underline{C}},\;\; C_3 = 2(2\sqrt{2}+4)^2,
\end{equation}
where $\underline{C}$ is as in condition (A1).
\vspace{0.1in}
\\
{\textbf{Step 3.1:}} For ease of presentation, we take $\lambda_T \leq \delta_T/3$; for more information on the general case of $\lambda_T \leq \delta_T/m$, for an $m>1$, see Remark 1 in the online supplementary material of \cite{anastasiou2019detecting}. Allow us now $\forall j \in \left\lbrace 1, \ldots, N\right\rbrace$, to define the intervals 
\begin{align}
\label{isolating_intervals}
\nonumber & I_{j}^R = \left[r_j + \frac{\delta_T}{3}, r_j + 2\frac{\delta_T}{3}\right)\\
&I_{j}^L = \left(r_j - 2\frac{\delta_T}{3}, r_j - \frac{\delta_T}{3}\right].
\end{align}
It is apparent that in order for $I_j^R$ and $I_j^L$ to have at least one point, then we actually implicitly require that $\delta_T > 3$, which is the case for sufficiently large $T$. Since the length of the intervals in \eqref{isolating_intervals} is equal to $\delta_T/3$ and $\lambda_T \leq \delta_T/3$, then MID ensures that for $K=\left\lceil T/\lambda_T  \right\rceil$ and $k,m \in \left\lbrace 1,\ldots,K\right\rbrace$, there exists at least one $c_k^r = k\lambda_T$ and at least one $c_m^l = T-m\lambda_T+1$ that are in $I_j^R$ and $I_j^L$, respectively, $\forall j =1,\ldots,N$.

At the beginning of our algorithm, $s=1$, $e=T$ (see, for example, Figure \ref{fig:steps}) and depending on whether $r_1 \leq T - r_N$ then $r_1$ or $r_N$ will get isolated in a right- or left-expanding interval, respectively. W.l.o.g., assume that $r_1 \leq T - r_N$ and, as already mentioned, our algorithm (which can be seen as an extension of the ID methodology in \cite{anastasiou2019detecting}) naturally ensures that $\exists k \in \left\lbrace 1,\ldots, K\right\rbrace$ such that $c_k^r \in I_1^R$. We highlight that there is no other change-point in $[1,c_k^r]$ apart from $r_1$. With $C_{s,e}^{b}$ and $D_{s,e}^b$ as in \eqref{set_A_T}, we will show that for $\tilde{b} = {\rm argmax}_{1\leq t < c_k^r}C_{1,c_k^r}^{t}$, then $C_{1,c_k^r}^{\tilde{b}} > \zeta_{T,d}$. Using \eqref{set_A_T}, we have that
\begin{align}
\label{thresholdpassing_firststep}
\nonumber & C_{1, c_k^r}^{\tilde{b}} \geq C_{1,c_k^r}^{r_1} \geq D_{1,c_k^r}^{r_1} - \sqrt{8\log\left(Td^{\frac{1}{4}}\right)}\\ 
\nonumber & = \sqrt{\frac{r_1(c_k^r - r_1)}{c_k^r}}\max\limits_{1 \leq j \leq d} \left\lbrace\Delta_1^{j}\right\rbrace- \sqrt{8\log\left(Td^{\frac{1}{4}}\right)}\\
& \geq \sqrt{\frac{\min\left\lbrace c_{k}^r - r_1, r_1\right\rbrace}{2}}\max\limits_{1 \leq j \leq d} \left\lbrace\Delta_1^{j}\right\rbrace- \sqrt{8\log\left(Td^{\frac{1}{4}}\right)},
\end{align}
with $\Delta_j^{q}$ defined in \eqref{general_notation}. For $\delta_T$ as in \eqref{general_notation}, we notice that $r_1 \geq \delta_T$. Furthermore, since $c_k^r \in I_1^R$ as in \eqref{isolating_intervals}, then $c_k^r - r_1 \geq \delta_T/3$, meaning that $\min\left\lbrace c_k^r - r_1, r_1 \right\rbrace \geq \frac{\delta_T}{3}$. Therefore, using assumption (A1),
\begin{align}
\label{proof_exceed}
\nonumber C_{1, c_k^r}^{\tilde{b}} & \geq \sqrt{\frac{\delta_T}{6}}\max\limits_{1 \leq j \leq d} \left\lbrace\Delta_1^{j}\right\rbrace - \sqrt{8\log\left(Td^{\frac{1}{4}}\right)}\\
\nonumber & \geq \sqrt{\frac{\delta_T}{6}}\underline{f}_T - \sqrt{8\log\left(Td^{\frac{1}{4}}\right)}\\
\nonumber & = \left(\frac{1}{\sqrt{6}} - \frac{\sqrt{8\log\left(Td^{1/4}\right)}}{\sqrt{\delta_T}\underline{f}_T}\right)\sqrt{\delta_T}\underline{f}_T\\
\nonumber & \geq \left(\frac{1}{\sqrt{6}} - \frac{\sqrt{8\log\left(Td^{1/4}\right)}}{\underline{C}\sqrt{\log\left(Td^{1/4}\right)}}\right)\sqrt{\delta_T}\underline{f}_T\\
& = \left(\frac{1}{\sqrt{6}} - \frac{2\sqrt{2}}{\underline{C}}\right)\sqrt{\delta_T}\underline{f}_T = C_2\sqrt{\delta_T}\underline{f}_T > \zeta_{T,d}.
\end{align}
Therefore, there will be an interval of the form $[1,c_{k}^r]$, with $c_{k}^r > r_1$, such that $[1,c_{k}^r]$ contains only $r_1$ and $\max_{1\leq b < c_{k}^r}\left\lbrace C_{1,c_{k}^r}^{b}\right\rbrace > \zeta_{T,d}$. Let us, for $k^* \in \left\lbrace 1,\ldots, K \right\rbrace$, to denote by $c_{k^*}^r \leq c_{k}^r$ the first right-expanding point where this happens and let $b_1 = {\rm argmax}_{1\leq t < c_{k^*}^r}\left\lbrace C_{1,c_{k^*}^r}^t\right\rbrace$ with $\left\lbrace C_{1,c_{k^*}^r}^{b_1}\right\rbrace > \zeta_{T,d}$. In addition, for ease of presentation, allow us to denote by $q_1:= {\rm argmax}_{j=1,\ldots,d}\left|\tilde{X}_{1,c_{k^*}^r}^{b_1,j}\right|$; this is the univariate component data sequence where the contrast function value got maximised at location $b_1$. We will find $\gamma_T>0$ such that for any $b^* \in \left\lbrace 1,\ldots,c_{k^*}^r - 1 \right\rbrace$ with $\left|b^*-r_1\right|\left(\Delta_1^{q_1}\right)^2 > \gamma_T$, it holds that
\begin{equation}
\label{relationship_contradiction}
\left(\tilde{X}_{1,c_{k^*}^r}^{r_1,q_1}\right)^2 > \left(\tilde{X}_{1,c_{k^*}^r}^{b^*,q_1}\right)^2.
\end{equation}
Proving \eqref{relationship_contradiction} and using the definition of $b_1$ we can conclude that $|b_1-r_1|\left(\Delta_1^{q_1}\right)^2 \leq \gamma_T$. Now, since in our model, $X_{t,j} = f_{t,j} + \epsilon_{t,j}$, then \eqref{relationship_contradiction} can be expressed as
\begin{align}
\label{relationship_contradiction2}
\nonumber & \left(\tilde{f}_{1,c_{k^*}^r}^{r_1,q_1}\right)^2 - \left(\tilde{f}_{1,c_{k^*}^r}^{b^*,q_1}\right)^2\\
\nonumber & > \left(\tilde{\epsilon}_{1,c_{k^*}^r}^{b^*,q_1}\right)^2 - \left(\tilde{\epsilon}_{1,c_{k^*}^r}^{r_1,q_1}\right)^2\\
&\; + 2\left\langle\boldsymbol{\psi_{1,c_{k^*}^r}^{b^*}}\langle\boldsymbol{f_{q_1}},\boldsymbol{\psi_{1,c_{k^*}^r}^{b^*}}\rangle - \boldsymbol{\psi_{1,c_{k^*}^r}^{r_1}}\langle\boldsymbol{f_{q_1}},\boldsymbol{\psi_{1,c_{k^*}^r}^{r_1}}\rangle,\boldsymbol{\epsilon_{q_1}}\right\rangle.
\end{align}
W.l.o.g. assume that $b^* \geq r_1$ and a similar approach as below holds when $b^*<r_1$. Using Lemma 4 in the online supplementary material of \cite{NOT}, gives for the left-hand side of the inequality in \eqref{relationship_contradiction2} that
\begin{equation}
\label{Lambda_mean}
\left(\tilde{f}_{1,c_{k^*}^r}^{r_1,q_1}\right)^2 - \left(\tilde{f}_{1,c_{k^*}^r}^{b^*,q_1}\right)^2 = \frac{\left|b^* - r_1\right|r_1}{\left|b^* - r_1\right| + r_1}\left(\Delta_1^{q_1}\right)^2 := \Lambda.
\end{equation}
For the terms on the right-hand side of \eqref{relationship_contradiction2}, since we are working under $A_T^*$ as in \eqref{set_A_T}, we obtain that
\begin{align}
\nonumber \left(\tilde{\epsilon}_{1,c_{k^*}^r}^{b^*,q_1}\right)^2 - \left(\tilde{\epsilon}_{1,c_{k^*}^r}^{r_1,q_1}\right)^2 & \leq \left(\max_{s \leq b < e}\left\lbrace\max_{1\leq j \leq d}\left|\tilde{\epsilon}_{s,e}^{b,j}\right|\right\rbrace\right)^2\\
\nonumber & \qquad - \left(\tilde{\epsilon}_{1,c_{k^*}^r}^{r_1,q_1}\right)^2\\
\nonumber & \leq \left(\max_{
s \leq b < e}\left\lbrace\max_{1\leq j \leq d}\left|\tilde{\epsilon}_{s,e}^{b,j}\right|\right\rbrace\right)^2\\
\nonumber & \leq 8\log\left(Td^{\frac{1}{4}}\right),
\end{align}
while from \eqref{set_B_T} and Lemma 4 from the online supplementary material of \cite{NOT},
\begin{align}
\nonumber & 2\left\langle\boldsymbol{\psi_{1,c_{k^*}^r}^{b^*}}\langle\boldsymbol{f_{q_1}},\boldsymbol{\psi_{1,c_{k^*}^r}^{b^*}}\rangle - \boldsymbol{\psi_{1,c_{k^*}^r}^{r_1}}\langle\boldsymbol{f_{q_1}},\boldsymbol{\psi_{1,c_{k^*}^r}^{r_1}}\rangle,\boldsymbol{\epsilon_{q_1}}\right\rangle\\
\nonumber & \leq 2\|\boldsymbol{\psi_{1,c_{k^*}^r}^{b^*}}<\boldsymbol{f_{q_1}},\boldsymbol{\psi_{1,c_{k^*}^r}^{b^*}}> - \boldsymbol{\psi_{1,c_{k^*}^r}^{r_1}}<\boldsymbol{f_{q_1}},\boldsymbol{\psi_{1,c_{k^*}^r}^{r_1}}>\|_{2}\\
\nonumber & \qquad \times\sqrt{8\log \left(Td^{\frac{1}{4}}\right)}\\
\nonumber & = 2\sqrt{8\Lambda\log \left(Td^{\frac{1}{4}}\right)}.
\end{align}
Therefore, \eqref{relationship_contradiction2} is satisfied if the stronger inequality $\Lambda > 8\log\left(Td^{1/4}\right) + 2\sqrt{8\Lambda\log\left(Td^{1/4}\right)}$ is satisfied, which has solution
\begin{equation}
\nonumber \Lambda > (2\sqrt{2}+4)^2\log\left(Td^{\frac{1}{4}}\right).
\end{equation}
From \eqref{Lambda_mean} and since 
\begin{equation}
\nonumber \frac{\left|b^*-r_1\right|r_1}{\left|b^*-r_1\right| + r_1} \geq \frac{1}{2}\min\left\lbrace\left|b^*-r_1\right|,r_1\right\rbrace,
\end{equation}
we deduce that \eqref{relationship_contradiction} is implied by
\begin{align}
\label{relationship_contradiction3}
\nonumber \min\left\lbrace\left|b^*-r_1\right|,r_1\right\rbrace & > \frac{2(2\sqrt{2}+4)^2\log\left( Td^{\frac{1}{4}}\right)}{\left(\Delta_1^{q_1}\right)^2}\\
& = \frac{C_3\log\left(Td^{\frac{1}{4}}\right)}{\left(\Delta_1^{q_1}\right)^2}.
\end{align}
However,
\begin{equation}
\label{relationship_contradiction4}
\min\left\lbrace r_1,c_{k^*}^r-r_1 \right\rbrace > C_3\frac{\log\left(Td^{\frac{1}{4}}\right)}{\left(\Delta_1^{q_1}\right)^2}
\end{equation}
and this is because if we assume that $\min \left\lbrace r_1,c_{k^*}^r-r_1 \right\rbrace \leq C_3\log\left(Td^{\frac{1}{4}}\right)/\left(\Delta_1^{q_1}\right)^2$, then
\begin{align}
\nonumber & C_{1,c_{k^*}^r}^{b_1} = \left|\tilde{X}_{1,c_{k^*}^r}^{b_1,q_1}\right| = \left|\tilde{X}_{1,c_{k^*}^r}^{b_1,q_1} - \tilde{f}_{1,c_{k^*}^r}^{b_1,q_1} + \tilde{f}_{1,c_{k^*}^r}^{b_1,q_1}\right|\\
\nonumber & \leq \sqrt{8\log\left(Td^{\frac{1}{4}}\right)} + \left|\tilde{f}_{1,c_{k^*}^r}^{b_1,q_1}\right| \leq \sqrt{8\log\left(Td^{\frac{1}{4}}\right)} + \left|\tilde{f}_{1,c_{k^*}^r}^{r_1,q_1}\right|\\
\nonumber & = \sqrt{8\log\left(Td^{\frac{1}{4}}\right)} + \sqrt{\frac{r_1(c_{k^*}^{r} - r_1)}{c_{k^*}^{r}}}\Delta_{1}^{q_1}\\
\nonumber & \leq \sqrt{8\log\left(Td^{\frac{1}{4}}\right)} + \sqrt{\min \left\lbrace c_{k^*}^r - r_1,r_1 \right\rbrace}\Delta_1^{q_1}\\
\nonumber & \leq \left(\sqrt{C_3} + \sqrt{8}\right)\sqrt{\log\left(Td^{\frac{1}{4}}\right)} = C_1\sqrt{\log\left(Td^{\frac{1}{4}}\right)}\\
\nonumber & \leq \zeta_{T,d}.
\end{align}
This comes to a contradiction to $C_{1,c_{k^*}^r}^{b_1} > \zeta_{T,d}$, which has already been proven in \eqref{proof_exceed}. Therefore, \eqref{relationship_contradiction4} holds and \eqref{relationship_contradiction3} is restricted to $\left|b^* - r_1\right|\left(\Delta_1^{q_1}\right)^2 > C_3\log T$, which implies \eqref{relationship_contradiction}. Therefore, necessarily
\begin{equation}
\label{distance1}
\left|b_1 - r_1\right|\left(\Delta_1^{q_1}\right)^2 \leq C_3 \log\left(Td^{\frac{1}{4}}\right).
\end{equation}
So far, for $\lambda_T \leq \delta_T/3$ we have proven that working under the assumption that $A_T^*$ (which implies that $A_T$ also holds) and $B_T$ hold, there will be an interval $[1,c_{k^*}^r]$, with $C_{1,c_{k^*}^r}^{b_1} >\zeta_{T,d}$, where $b_1=\underset{1\leq t < c_{k^*}^r}{{\rm argmax}}C_{1,c_{k^*}^r}^{t}$ is an estimation of $r_1$ that satisfies \eqref{distance1}.
\\
{\textbf{Step 3.2:}} After detecting the first change-point, MID follows the same process as in Step 3.1 but now in the set $[c_{k^*}^r,T]$, which contains the change-points $r_2, \ldots, r_N$. We do not check for possible change-points the interval $[b_1+1, c_{k^*}^r)$. However, this does not create any issues because:
\begin{itemize}
{\small{
\item[M.1] There is no change-point in $[b_1 + 1,c_{k^*}^r)$, apart from maybe the already detected $r_1$;
\item[M.2] $c_{k^*}^r$ is at a location which allows for detection of $r_2$.}}
\end{itemize}
{\textbf{For M.1:}} We will split the explanation into two cases with respect to the location of $b_1$.\\
{\textbf{Case 1:}} $b_1 < r_1 < c_{k^*}^r$. Using \eqref{distance1} and with
\begin{equation}
\label{condition_delta1}
\delta_T > 3C_3\frac{\log\left(Td^{\frac{1}{4}}\right)}{\left(\Delta_1^{q_1}\right)^2},
\end{equation}then since $c_{k}^r \in I_1^R$, we have that
\begin{align}
\nonumber c_{k^*}^r - b_1 & \leq c_{k}^r - b_1 = c_{k}^r - r_1 + r_1 - b_1\\
\nonumber & < 2\frac{\delta_T}{3} + r_1 - b_1\\
\nonumber &\leq 2\frac{\delta_T}{3} + \frac{C_3\log\left(Td^{\frac{1}{4}}\right)}{\left(\Delta_1^{q_1}\right)^2} < \delta_T.
\end{align}
Since $r_2 - r_1 \geq \delta_T$ and $r_1$ is already in $[b_1+1,c_{k^*}^r)$, then there is no other change-point in $[b_1+1,c_{k^*}^r)$ apart from $r_1$.\\
{\textbf{Case 2:}} $r_1 \leq b_1 < c_{k^*}^r$. By construction $c_{k^*}^r - r_1 < 2\delta_T/3$, which means that apart from $r_1$ there is no other change-point in $[r_1,c_{k^*}^r)$. With $r_1 \leq b_1$, then $[b_1+1,c_{k^*}^r)$ does not have any change-point.

Cases 1 and 2 above show that no matter the location of $b_1$, there is no change-point in $[b_1+1,c_{k^*}^r)$ other than possibly the previously detected $r_1$. Similarly to the approach in Step 3.1, our method applied now in $[s,e] = [c_{k^*}^r,T]$,  will first isolate, and then detect, $r_2$ or $r_N$ depending on whether $r_2 - c_{k^*}^r$ is smaller or larger than $T-r_N$.
If $T-r_N < r_2 - c_{k^*}^r$ then $r_N$ will get isolated first in a left-expanding interval and the procedure to show its detection is exactly the same as for the detection of $r_1$ in Step 3.1. Therefore, for the sake of showing M.2 let us assume that $r_2 - c_{k^*}^r \leq T-r_N$.
\vspace{0.1in}
\\
{\textbf{For M.2:}} By construction, there exists a right expanding point $c_{k_2}^r \in I_2^R$, with $I_{j}^R$ defined in \eqref{isolating_intervals}. We will show that $r_2$ gets detected in $[c_{k^*}^r,c_{k^*_2}^r]$, for $k_2^* \leq k_2$ and its detection is $b_2 = {\rm argmax}_{c_{k^*}^r \leq t < c_{k^*_2}^r}C_{c_{k^*}^r,c_{k^*_2}^r}^t$, which satisfies $\left|b_2 - r_2\right|\left(\Delta_2^{q_2}\right)^2 \leq C_3\log \left(Td^{1/4}\right)$, where $q_2:= {\rm argmax}_{j=1,2,\ldots,d}\left|\tilde{X}_{c_{k^*}^r, c_{k^*_2}^r}^{b_2,j}\right|$. Following similar steps as in \eqref{proof_exceed}, we have that 
\begin{equation}
\nonumber C_{c_{k^*}^r,c_{k_2^*}^r}^{b_2} >\zeta_{T,d}.
\end{equation}
We will now show that $\left|b_2 - r_2\right|\left(\Delta_2^{q_2}\right)^2 \leq C_3\log \left(Td^{1/4}\right)$. Following exactly the same process as in Step 3.1 and assuming now w.l.o.g. that $b_2 < r_2$, we have that for $b^* \in \left\lbrace c_{k^*}^r,\ldots,c_{k^*_2}^r - 1\right\rbrace$, 
\begin{equation}
\label{r2_relationship_contradiction}
\left(\tilde{X}_{c_{k^*}^r,c_{k^*_2}^r}^{r_2,q_2}\right)^2 > \left(\tilde{X}_{c_{k^*}^r,c_{k^*_2}^r}^{b^*,q_2}\right)^2
\end{equation}
is implied by $\min \left\lbrace \left|b^* - r_2\right|,c_{k^*_2}^r - r_2 \right\rbrace > C_3\log T/\left(\Delta_2^f\right)^2$. In the same way as in Step 3.1 and by contradiction we can show that $\min \left\lbrace c_{k^*_2}^r - r_2, r_2 - c_{k^*}^r + 1\right\rbrace > C_3\log T/\left(\Delta_2^f\right)^2$ and \eqref{r2_relationship_contradiction} is implied by $\left|b^* - r_2\right|\left(\Delta_2^{q_2}\right)^2 > C_3\log\left(Td^{1/4}\right)$. Therefore $\left|b_2 - r_2\right|\left(\Delta_2^{q_2}\right)^2 > C_3 \log \left(Td^{1/4}\right)$ would mean that $\left|\tilde{X}_{c_{k^*}^r,c_{k^*_2}^r}^{r_2,q_2}\right| > \left|\tilde{X}_{c_{k^*}^r,c_{k^*_2}^r}^{b_2,r_2}\right|$, which is not true by the definition of $b_2$. Having said this, we conclude that $\left|b_2 - r_2\right|\left(\Delta_2^{q_2}\right)^2 \leq C_3\log \left(Td^{1/4}\right)$. Having detected $r_2$, then our algorithm will proceed in the interval $[s,e]=[c_{k^*_2}^r, T]$ and all the change-points will get detected one by one since Step 3.2 will be applicable as long as there are undetected change-points in $[s,e]$.

Denoting by $\hat{r}_j$ the estimation of $r_j$ as we did in the statement of the theorem and for $[s^*,e^*]$ being the interval where the isolation and detection of $r_j$ occurs (in the way that we explained in Steps 3.1 and 3.2) allow us to denote by $q_j:= {\rm argmax}_{m=1,2,\ldots,d}\left|\tilde{X}_{s^*, e^*}^{\hat{r}_j,m}\right|$. Then, Steps 3.1 and 3.2 have shown that MID will detect all the change-points one by one and $\left|\hat{r}_j - r_j\right|\left(\Delta_j^{q_j}\right)^2 \leq C_3\log \left(Td^{1/4}\right), \quad \forall j \in \left\lbrace 1,2,\ldots,N \right\rbrace$.
\\
{\textbf{Step 4:}} The arguments given in Steps 1-3 hold in $A_T^* \cap B_T$. At the beginning of the algorithm, $s=1, e=T$ and for $N\geq 1$, there exist $k_1\in \left\lbrace 1,\ldots, K \right\rbrace$ such that $s_{k_1} = s, e_{k_1} \in I_1^R$ and $k_2\in \left\lbrace 1,\ldots, K \right\rbrace$ such that $s_{k_2} \in I_N^L, e_{k_2} = e$. As in our previous steps, w.l.o.g. assume that $r_1 \leq T - r_N$ and $r_1$ gets isolated and detected first in an interval $[s,c_{k^*}^r]$, where $c_{k^*}^r$ is less than or equal to $e_{k_1}$. Then, $\hat{r}_1 = {\rm argmax}_{s \leq t < c_{k^*}^r}C_{s,c_{k^*}^r}^t$ is the estimated location for $r_1$ and $\left|r_1-\hat{r}_1\right|\left(\Delta_1^{q_1}\right)^2\leq C_3\log\left(Td^{1/4}\right)$. After this, the method continues in $[c_{k^*}^r,T]$ and keeps detecting all the change-points as explained in Step 3. There will not be any double detection issues because naturally, at each step of the algorithm, the new interval $[s,e]$ does not include any previously detected change-points. Once all the change-points have been detected one by one, then $[s,e]$ will contain no other change-points. Our method will of course keep checking for possible change-points in right- and left-expanding intervals denoted by $[s^*,e^*]$. However, MID will not detect anything in $[s^*,e^*]$ because $\forall b \in [s^*,e^*)$,
\begin{align}
\nonumber C_{s^*,e^*}^{b} & \leq D_{s^*,e^*}^{b} + \sqrt{8\log \left(Td^{\frac{1}{4}}\right)} = \sqrt{8\log \left(Td^{\frac{1}{4}}\right)}\\
\nonumber & < C_1\sqrt{\log \left(Td^{\frac{1}{4}}\right)}\leq \zeta_{T,d}.
\end{align}
After not detecting anything in all intervals of the above form, then the algorithm concludes that there are not any more change-points to be detected and stops.
\end{proof}
\section{Proof of Theorem \ref{theorem_consistency_S2}}
\label{app:proofs}
\vspace{0.05in}
Before proving the result, we introduce some useful notation with
\begin{align}
\label{G_H_notation}
\nonumber G^{b}_{s,e} & := \max\limits_{1 \leq j \leq d}\left\lbrace C_{s,e}^b(\boldsymbol{X_j})\right\rbrace\\
H_{s,e}^{b} & := \max\limits_{1 \leq j \leq d}\left\lbrace C_{s,e}^b(\boldsymbol{f_j})\right\rbrace,
\end{align}
where $C_{s,e}^b(\cdot)$ is as in \eqref{contrast_linear} of the main paper. For ease of presentation, we define the sets
\begin{align}
\label{set_A_T_linear}
\nonumber \tilde{A}_T^* & =\bigg\{\max_{1\leq s \leq b < e\leq T}\left\lbrace\max_{1 \leq j \leq d}\left|C_{s,e}^b(\boldsymbol{X_j}) -C_{s,e}^b(\boldsymbol{f_j})\right|\right\rbrace\leq \sqrt{8\operatorname{log}\left(Td^{\frac{1}{4}}\right)}\bigg\},\\
\tilde{A}_T & =\bigg\{\max_{1 \leq s \leq b < e \leq T}\left|G^{b}_{s,e}-H_{s,e}^{b}\right|\leq \sqrt{8\operatorname{log}\left(Td^{\frac{1}{4}}\right)}\bigg\},\\
\nonumber \tilde{B}_T &= \left\{\max _{\substack{j=1, \ldots, N\\k = 1,\ldots, d}} \max_{\substack{
r_{j-1}<s \leq r_{j}\\
r_{j} < e \leq r_{j+1} \\
s \leq b <e}}\left|\tilde{A}_{s,e}^{b}(k,r_j)\right| \leq \sqrt{8\operatorname{log}\left(Td^{\frac{1}{4}}\right)}\right\},
\end{align}
where for $\boldsymbol{\phi_{s, e}^{b}}$ as in \eqref{contrast_vectorCPLM} of the main paper,
\begin{equation}
\label{A_seb_linear}
    \tilde{A}_{s,e}^{b}(k,r_j) := \frac{\left\langle\boldsymbol{\phi_{s, e}^{b}}\left\langle \boldsymbol{f_{k}}, \boldsymbol{\phi_{s, e}^{b}}\right\rangle-\boldsymbol{\phi_{s,e}^{r_{j}}}\left\langle\boldsymbol{f_{k}},\boldsymbol{\phi_{s,e}^{r_{j}}}\right\rangle,\boldsymbol{\epsilon}\right\rangle|}{\left\|\boldsymbol{\phi_{s, e}^{b}}\left\langle \boldsymbol{f_{k}}, \boldsymbol{\phi_{s,e}^{b}}\right\rangle-\boldsymbol{\phi_{s,e}^{r_{j}}}\left\langle\boldsymbol{f_{k}},\boldsymbol{\phi_{s,e}^{r_{j}}}\right\rangle\right\|_{2}}.
\end{equation}
\begin{proof}
We will prove the more specific result
\begin{equation}
\label{mainresult_theorem3}
\Prob\left(\hat{N} = N, \max_{j=1,\ldots,N}\left(\left|\hat{r}_j - r_j\right|\left(\Delta_j^{q_j}\right)^{\frac{2}{3}}\right) \leq C_3(\log \left(Td^{1/4}\right))^\frac{1}{3}\right) \leq 1 - \frac{1}{6\sqrt{\pi}T},
\end{equation}
which implies the result in \eqref{mainresult_theorem_S2} of the main paper.\\
{\textbf{Steps 1 and 2}:}  Due to the fact that $\Prob\left(\tilde{A}_T\right) \geq \Prob\left(\tilde{A}_T^*\right)$, then arguments as those used in Steps 1 and 2 of the proof of Theorem \ref{theorem_consistency_S1} can be employed here as well in order to show that 
\begin{align}
\label{result1_linear}
\nonumber & \Prob\left(\tilde{A}_T\right) \geq \Prob\left(\tilde{A}_T^*\right) \geq 1 - \frac{1}{12\sqrt{\pi}T}\\
& \Prob\left(\tilde{B}_T\right) \geq 1 - \frac{1}{12\sqrt{\pi}T}.
\end{align}
Therefore, $\Prob\left(\tilde{A}_T \cap \tilde{B}_T\right) \geq 1 - 1/(6\sqrt{\pi}T)$.
\vspace{0.1in}
\\
{\textbf{Step 3:}} This is the main part of our proof, where we explain in detail how to get the result in \eqref{mainresult_theorem3}. From now on, we assume that $\tilde{A}_T^*$ (therefore, $\tilde{A}_T$ as well) and $\tilde{B}_T$ both hold. The constants we use are
\begin{equation}
\nonumber C_1 = \sqrt{\frac{2}{3}}C_3^{\frac{3}{2}} + \sqrt{8},\;\; C_2 = \frac{1}{3\sqrt{72}} - \frac{2\sqrt{2}}{C^*},\;\; C_3 = 63^{\frac{1}{3}}(2\sqrt{2}+4)^{\frac{2}{3}},
\end{equation}
where $C^*$ is as in assumption (A2) of the main paper.
\vspace{0.1in}
\\
{\textbf{Step 3.1:}} First, $\forall j \in \left\lbrace 1, \ldots, N\right\rbrace$, we define $I_{j}^R$ and $I_{j}^L$ as in \eqref{isolating_intervals}. At the beginning of our algorithm, $s=1$, $e=T$ and depending on whether $r_1 \leq T - r_N$, then $r_1$ or $r_N$ will get isolated first in a right- or left-expanding interval, respectively. W.l.o.g., assume that $r_1 \leq T - r_N$. Because of the structure of the ID methodology developed in \cite{anastasiou2019detecting} and partly employed in our paper, it is ensured that $\exists \tilde{k} \in \left\lbrace 1,\ldots, K\right\rbrace$ such that $c_{\tilde{k}}^r = \tilde{k}\lambda_T \in I_1^R$ and there is no other change-point in $[1,c_{\tilde{k}}^r]$ apart from $r_1$. With, our aim now is to show that for $\tilde{b}_1 = {\rm argmax}_{1 < t < c_{\tilde{k}}^r}G_{1,c_{\tilde{k}}^r}^t$, then $G_{1,c_{\tilde{k}}^r}^{\tilde{b}_1} > \zeta_{T,d}$. Using \eqref{set_A_T_linear}, we obtain that
\begin{align}
\label{thresholdpassing_firststep_trend}
\nonumber G_{1,c_{\tilde{k}}^r}^{\tilde{b}_1} & \geq G_{1,c_{\tilde{k}}^r}^{r_1} \geq H_{1,c_{\tilde{k}}^r}^{r_1} - \sqrt{8 \log \left(Td^{\frac{1}{4}}\right)} =  \max\limits_{1 \leq j \leq d}\left\lbrace C_{1,c_{\tilde{k}}^r}^{r_1}(\boldsymbol{f_j})\right\rbrace - \sqrt{8\log\left(Td^{\frac{1}{4}}\right)}\\
\nonumber & \geq \max\limits_{1 \leq j \leq d}\left\lbrace\frac{1}{\sqrt{24}}\left(\min\left\lbrace r_1-1, c_{\tilde{k}}^r - r_1\right\rbrace \right)^{3/2}\Delta_1^j\right\rbrace - \sqrt{8\log\left(Td^{\frac{1}{4}}\right)}\\
& = \frac{1}{\sqrt{24}}\left(\min\left\lbrace r_1-1, c_{\tilde{k}}^r - r_1\right\rbrace \right)^{3/2}\max\limits_{1 \leq j \leq d}\left\lbrace\Delta_1^j\right\rbrace - \sqrt{8\log\left(Td^{\frac{1}{4}}\right)},
\end{align} 
where the third inequality is due to the result of Lemma 5 in the online supplement of \cite{NOT}. Now, $r_1 -1 = r_1 - r_0 - 1 \geq \delta_T -1 > \delta_T/3$, because in the case of the continuous piecewise-linear signals of Scenario (S2) we necessarily have that $\delta_T \geq 2$; otherwise change-point identifiability would not be possible. In addition, since $c_{\tilde{k}}^r \in I_1^R$, then $c_{\tilde{k}}^r - r_1 \geq \delta_T/3$, meaning that
\begin{equation}
\label{mindistance1_trend}
\min\left\lbrace c_{\tilde{k}}^r - r_1, r_1 -1 \right\rbrace \geq \frac{\delta_T}{3}.
\end{equation}
The result in \eqref{thresholdpassing_firststep_trend}, Assumption (A2) in Section \ref{subsec:cplm} of the main paper, and \eqref{mindistance1_trend} yield
\begin{align}
\label{thresholdpassing_trend}
\nonumber G_{1,c_{\tilde{k}}^r}^{\tilde{b}_1} & \geq \frac{1}{\sqrt{24}}\left(\frac{\delta_T}{3}\right)^{3/2}\max\limits_{1 \leq j \leq d}\left\lbrace\Delta_1^j\right\rbrace - \sqrt{8\log\left(Td^{\frac{1}{4}}\right)}\\
\nonumber & \geq \frac{1}{\sqrt{24}}\left(\frac{\delta_T}{3}\right)^{3/2}\underline{f}_T - \sqrt{8\log\left(Td^{\frac{1}{4}}\right)}\\
\nonumber & = \delta_T^{3/2}\underline{f}_T\left(\frac{1}{3\sqrt{72}} - \frac{\sqrt{8\log\left(Td^{\frac{1}{4}}\right)}}{\delta_T^{3/2}\underline{f}_T}\right) \geq \left(\frac{1}{3\sqrt{72}}- \frac{2\sqrt{2}}{C^*}\right)\delta_T^{3/2}\underline{f}_T\\
& = C_2\delta_T^{3/2}\underline{f}_T > \zeta_{T,d}.
\end{align}
The result in \eqref{thresholdpassing_trend} above proves that there will be an interval of the form $[1,c_{\tilde{k}}^r]$, with $c_{\tilde{k}}^r > r_1$, such that $[1,c_{\tilde{k}}^r]$ contains only $r_1$ and $\max_{1\leq b < c_{\tilde{k}}^r}G_{1,c_{\tilde{k}}^r}^{b} > \zeta_{T,d}$. Let us, for $k^* \in \left\lbrace 1,\ldots, K \right\rbrace$, denote by $c_{k^*}^r \leq c_{\tilde{k}}^r$ the first right-expanding point where this happens and let $b_1 = {\rm argmax}_{1\leq t < c_{k^*}^r}G_{1,c_{k^*}^r}^t$ with $G_{1,c_{k^*}^r}^{b_1} > \zeta_{T,d}$. Note that $b_1$ can not be an estimation of any other change-point as $[1,c_{k^*}^r]$ includes only $r_1$.

Allow us, for ease of presentation, to denote by $q_1:= {\rm argmax}_{j=1,\ldots,d}\left\lbrace C_{1,c_{k^*}^r}^{b_1}(\boldsymbol{X_j})\right|$ and $q_1$ is basically the location index of the univariate component data sequence where the contrast function value got maximised at point $b_1$. Apparently, it holds that
\begin{equation}
\nonumber G_{1,c_{k^*}^r}^{b_1} = \max\limits_{1 \leq j \leq d}\left\lbrace C_{1,c_{k^*}^r}^{b_1}(\boldsymbol{X_j})\right\rbrace = C_{1,c_{k^*}^r}^{b_1}(\boldsymbol{X_{q_1}}).
\end{equation}
Our aim now is to find $\tilde{\gamma}_T > 0$ such that for any $b^* \in \left\lbrace 1,\ldots,c_{k^*}^r - 1 \right\rbrace$ with $\left|b^*-r_1\right|\left(\Delta_1^{q_1}\right)^{2/3} > \tilde{\gamma}_T$, it holds that
\begin{equation}
\label{relationship_contradiction_trend}
\left(C_{1,c_{k^*}^r}^{r_1}(\boldsymbol{X_{q_1}})\right)^2 > \left(C_{1,c_{k^*}^r}^{b^*}(\boldsymbol{X_{q_1}})\right)^2.
\end{equation}
Proving \eqref{relationship_contradiction_trend} and using the definition of $b_1$, we can then conclude that $|b_1-r_1|\left(\Delta_1^{q_1}\right)^{2/3} \leq \tilde{\gamma}_T$. The inequality in \eqref{relationship_contradiction_trend} can be expressed as
\begin{align}
\label{relationship_contradiction2_trend}
\nonumber \left(C_{1,c_{k^*}^r}^{r_1}(\boldsymbol{f_{q_1}})\right)^2 - \left(C_{1,c_{k^*}^r}^{b^*}(\boldsymbol{f_{q_1}})\right)^2 & > \left(C_{1,c_{k^*}^r}^{b^*}(\boldsymbol{\epsilon_{q_1}})\right)^2 - \left(C_{1,c_{k^*}^r}^{r_1}(\boldsymbol{\epsilon_{q_1}})\right)^2\\
& + 2\left\langle\boldsymbol{\phi_{1,c_{k^*}^r}^{b^*}}\langle\boldsymbol{f_{q_1}},\boldsymbol{\phi_{1,c_{k^*}^r}^{b^*}}\rangle - \boldsymbol{\phi_{1,c_{k^*}^r}^{r_1}}\langle\boldsymbol{f_{q_1}},\boldsymbol{\phi_{1,c_{k^*}^r}^{r_1}}\rangle,\boldsymbol{\epsilon_{q_1}}\right\rangle.
\end{align}
W.l.o.g. assume that $b^* \geq r_1$ and a similar approach as below holds when $b^*<r_1$. We denote by
\begin{equation}
\nonumber \Lambda := \left(C_{1,c_{k^*}^r}^{r_1}(\boldsymbol{f_{q_1}})\right)^2 - \left(C_{1,c_{k^*}^r}^{b^*}(\boldsymbol{f_{q_1}})\right)^2
\end{equation}
and for the terms in the right-hand side of \eqref{relationship_contradiction2_trend}, we get that
\begin{align}
\nonumber \left(C_{1,c_{k^*}^r}^{b^*}(\boldsymbol{\epsilon_{q_1}})\right)^2 - \left(C_{1,c_{k^*}^r}^{r_1}(\boldsymbol{\epsilon_{q_1}})\right)^2 & \leq  \left(\max_{s,e,b:s\leq b <e}\left\lbrace\max_{1 \leq j \leq d}\left\lbrace C_{s,e}^{b}(\boldsymbol{\epsilon_j})\right\rbrace\right\rbrace\right)^2\\
\nonumber & \;\; - \left(C_{1,c_k^r}^{r_1}(\boldsymbol{\epsilon_{q_1}})\right)^2\\
\nonumber & \leq \left(\max_{s,e,b:s\leq b <e}\left\lbrace\max_{1 \leq j \leq d}\left\lbrace C_{s,e}^{b}(\boldsymbol{\epsilon_j})\right\rbrace\right\rbrace\right)^2\\ \nonumber & \leq 8\log \left(Td^{\frac{1}{4}}\right),
\end{align}
while from \eqref{set_A_T_linear} and Lemma 7 in the online supplement of \cite{NOT}, 
\begin{align}
\nonumber & 2\left\langle\boldsymbol{\phi_{1,c_{k^*}^r}^{b^*}}\langle\boldsymbol{f_{q_1}},\boldsymbol{\phi_{1,c_{k^*}^r}^{b^*}}\rangle - \boldsymbol{\phi_{1,c_{k^*}^r}^{r_1}}\langle\boldsymbol{f_{q_1}},\boldsymbol{\phi_{1,c_{k^*}^r}^{r_1}}\rangle,\boldsymbol{\epsilon_{q_1}}\right\rangle \\
\nonumber & \leq 2\|\boldsymbol{\phi_{1,c_{k^*}^r}^{b^*}}<\boldsymbol{f_{q_1}},\boldsymbol{\phi_{1,c_{k^*}^r}^{b^*}}> - \boldsymbol{\phi_{1,c_{k^*}^r}^{r_1}}<\boldsymbol{f_{q_1}},\boldsymbol{\phi_{1,c_{k^*}^r}^{r_1}}>\|_{2}\sqrt{8\log \left(Td^{\frac{1}{4}}\right)}\\
\nonumber & = 2\sqrt{8\Lambda\log\left(Td^{\frac{1}{4}}\right)}.
\end{align}
Therefore \eqref{relationship_contradiction2_trend} is satisfied if the stronger inequality $\Lambda > 8\log\left(Td^{1/4}\right) + 2\sqrt{8\Lambda\log\left(Td^{1/4}\right)}$ is satisfied, which has solution
\begin{equation}
\label{step_middle_trend}
\Lambda > (2\sqrt{2}+4)^2\log\left(Td^{\frac{1}{4}}\right).
\end{equation}
Using now Lemma 7 in the online supplement of \cite{NOT}, we have that \eqref{step_middle_trend} is implied by
\begin{align}
\label{relationship_contradiction3_trend}
\nonumber & \frac{1}{63}\left(\min\left\lbrace \left|r_1 - b^*\right|,r_1 - 1 \right\rbrace\right)^3\left(\Delta_1^{q_1}\right)^2 > \left(2\sqrt{2} + 4\right)^2\log\left(Td^{\frac{1}{4}}\right)\\
\nonumber & \Leftrightarrow \min\left\lbrace \left|r_1 - b^*\right|,r_1 - 1 \right\rbrace >  \frac{\left(63\log \left(Td^{\frac{1}{4}}\right)\right)^{1/3}(2\sqrt{2}+4)^{2/3}}{\left(\Delta_1^{q_1}\right)^{2/3}}\\
& \qquad\qquad\qquad\qquad\qquad\quad\;\; = \frac{C_3\left(\log \left(Td^{\frac{1}{4}}\right)\right)^{1/3}}{\left(\Delta_1^{q_1}\right)^{2/3}}.
\end{align}
However,
\begin{align}
\label{relationship_contradiction4_trend}
\min\left\lbrace r_1-1,c_{k^*}^r-r_1 \right\rbrace > 2^{1/3}C_3\frac{\left(\log \left(Td^{\frac{1}{4}}\right)\right)^{1/3}}{\left(\Delta_1^{q_1}\right)^{2/3}} - 1
\end{align}
and this is because if we assume that $$\min \left\lbrace r_1-1,c_{k^*}^r-r_1 \right\rbrace \leq 2^{1/3}C_3\left(\log \left(Td^{\frac{1}{4}}\right)\right)^{1/3}/\left(\Delta_1^{q_1}\right)^{2/3} - 1$$ then we obtain that
\begin{align}
\label{result_trend_proof_middle}
\nonumber G_{1,c_{k^*}^r}^{b_1} & = C_{1,c_{k^*}^r}^{b_1}(\boldsymbol{X_{q_1}})  \leq \left|C_{1,c_{k^*}^r}^{b_1}(\boldsymbol{X_{q_1}}) - C_{1,c_{k^*}^r}^{b_1}(\boldsymbol{f_{q_1}})\right| + C_{1,c_{k^*}^r}^{b_1}(\boldsymbol{f_{q_1}}) \\
\nonumber & \leq \sqrt{8\log \left(Td^{\frac{1}{4}}\right)} + C_{1,c_{k^*}^r}^{b_1}(\boldsymbol{f_{q_1}})\\
\nonumber & \leq \frac{1}{\sqrt{3}}\left(\min\left\lbrace r_1 -1, c_{k^*}^r - r_1 \right\rbrace + 1\right)^{3/2}\Delta_1^{q_1} + \sqrt{8\log\left(Td^{\frac{1}{4}}\right)}\\
\nonumber & \leq \frac{1}{\sqrt{3}}\left(2^{1/3}C_3\frac{\left(\log \left(Td^{\frac{1}{4}}\right)\right)^{1/3}}{\left(\Delta_1^{q_1}\right)^{2/3}}\right)^{3/2}\Delta_1^{q_1} + \sqrt{8\log\left(Td^{\frac{1}{4}}\right)}\\
\nonumber & = \sqrt{\frac{2}{3}}C_3^{3/2}\sqrt{\log \left(Td^{\frac{1}{4}}\right)} + \sqrt{8\log\left(Td^{\frac{1}{4}}\right)}\\
& = \left(\sqrt{\frac{2}{3}}C_3^{3/2} + \sqrt{8}\right)\sqrt{\log\left(Td^{\frac{1}{4}}\right)} = C_1\sqrt{\log\left(Td^{\frac{1}{4}}\right)} \leq \zeta_{T,d},
\end{align}
where the third inequality is a direct application of Lemma 5 in the online supplement of \cite{NOT}. The result in \eqref{result_trend_proof_middle} comes to a contradiction to $G_{1,c_{k^*}^r}^{b_1} > \zeta_{T,d}$ which has already been proven in \eqref{thresholdpassing_trend}. Therefore, \eqref{relationship_contradiction4_trend} holds and for $Td^{1/4}$ being sufficiently large,
\begin{equation}
\label{relationship_contradiction3_2_trend}
\min\left\lbrace r_1 -1,c_{k^*}^r - r_1 \right\rbrace > 2^{1/3}C_3\frac{\left(\log \left(Td^{\frac{1}{4}}\right)\right)^{1/3}}{\left(\Delta_1^{q_1}\right)^{2/3}} - 1> C_3\frac{\left(\log\left(Td^{\frac{1}{4}}\right)\right)^{1/3}}{\left(\Delta_1^{q_1}\right)^{2/3}}.
\end{equation}
From \eqref{relationship_contradiction3_2_trend} we deduce that \eqref{relationship_contradiction3_trend} is restricted to
\begin{equation}
\nonumber \left|r_1 - b^*\right| > C_3\frac{\left(\log \left(Td^{\frac{1}{4}}\right)\right)^{1/3}}{\left(\Delta_1^{q_1}\right)^{2/3}}, 
\end{equation}
which implies \eqref{relationship_contradiction_trend}. Therefore, necessarily,
\begin{equation}
\label{distance1_trend}
\left|b_1 - r_1\right|\left(\Delta_1^{q_1}\right)^{2/3} \leq C_3 \left(\log \left(Td^{\frac{1}{4}}\right)\right)^{1/3}.
\end{equation}
So far, we have proven that working under the sets $\tilde{A}_T^*$ and $\tilde{B}_T$, there will be an interval of the form $[1,c_{k^*}^r]$, with $G_{1,c_{k^*}^r}^{b_1} > \zeta_{T,d}$, where $b_1=\underset{1\leq t < c_{k^*}^r}{\rm argmax}\;G_{1,c_{k^*}^r}^{t}$ is an estimation of $r_1$ that satisfies \eqref{distance1_trend}.
\vspace{0.1in}
\\
{\textbf{Step 3.2:}}  After detecting the first change-point, MID follows the same process as in Step 3.1 in the set $[c_{k^*}^r,T]$, which contains $r_2, \ldots, r_N$. This means that we do not check for possible change-points in the interval $[b_1+1, c_{k^*}^r)$ and we need to prove that:
\begin{itemize}
\item[(M.1)] There is no other change-point in $[b_1 + 1,c_{k^*}^r)$, apart from possibly the already detected $r_1$;
\item[(M.2)] $c_{k^*}^r$ is at a location which allows for detection of $r_2$.
\end{itemize}
{\textbf{For (M.1):}} The approach is the same as the one in Step 3.2 in the proof of Theorem \ref{theorem_consistency_S1} and will not be repeated here.

Similarly to the approach in Step 3.1, our method applied now to $[c_{k^*}^r,T]$,  will first detect $r_2$ or $r_N$ depending on whether $r_2 - c_{k^*}^r$ is smaller or larger than $T-r_N$. If $T-r_N < r_2 - c_{k^*}^r$ then $r_N$ will get isolated first and the procedure to show its detection is exactly the same as in Step 3.1 where we explained the detection of $r_1$. Therefore, w.l.o.g. and also for the sake of showing (M.2) let us assume that $r_2 - c_{k^*}^r \leq T-r_N$.
\vspace{0.1in}
\\
{\textbf{For (M.2):}} Due to the isolation aspect of MID, it is certain that there exists a right expanding point $c_{k_2}^r$, such that $c_{k_2}^r \in I_2^R$, with $I_j^R$ being defined in \eqref{isolating_intervals} of the main paper. We will show that $r_2$ gets detected in $[c_{k^*}^r,c_{k^*_2}^r]$, for $k_2^* \leq k_2$ and its detection is $b_2 = {\rm argmax}_{c_{k^*}^r \leq t < c_{k^*_2}^r}G_{c_{k^*}^r,c_{k^*_2}^r}^t$, which satisfies $\left|b_2 - r_2\right|\left(\Delta_2^{q_2}\right)^{2/3} \leq C_3\left(\log \left(Td^{1/4}\right)\right)^{1/3}$, where $q_2:= {\rm argmax}_{j=1,\ldots,d}\left\lbrace C_{c_{k^*}^r, c_{k^*_2}^r}^{b_2}(\boldsymbol{X_j})\right\rbrace$. Using again Lemma 5 from the online supplement of \cite{NOT} and for $\tilde{b}_2 = {\rm argmax}_{c_{k^*}^r \leq t < c_{k_2}^r}G_{c_{k^*}^r,c_{k_2}^r}^t$, we have that
\begin{align}
\label{midstepforr2_trend}
\nonumber G_{c_{k^*}^r,c_{k_2}^r}^{\tilde{b}_2} & \geq G_{c_{k^*}^r,c_{k_2}^r}^{r_2} \geq H_{c_{k^*}^r,c_{k_2}^r}^{r_2} - \sqrt{8 \log \left(Td^{\frac{1}{4}}\right)}\\
& \geq \frac{1}{\sqrt{24}}\left(\min \left\lbrace r_2 - c_{k^*}^r, c_{k_2}^r-r_2 \right\rbrace\right)^{3/2}\underline{f}_T - \sqrt{8\log \left(Td^{\frac{1}{4}}\right)}.
\end{align}
By construction,
\begin{align}
\nonumber & c_{k_2}^r - r_2 \geq \frac{\delta_T}{3}\\
\nonumber & r_2 - c_{k^*}^r \geq r_2 - c_{k}^r = r_2 - r_1 - (c_k^r - r_1) \geq \delta_T - (c_k^r - r_1) > \delta_T - 2\frac{\delta_T}{3}  = \frac{\delta_T}{3},
\end{align}
which means that $\min\left\lbrace c_{k_2}^r-r_2,r_2-c_{k^*}^r \right\rbrace \geq \delta_T/3$. Therefore, continuing from \eqref{midstepforr2_trend} and using the exact same calculations as in \eqref{thresholdpassing_trend}, we have that
\begin{align}
\nonumber G_{c_{k^*}^r,c_{k_2^*}^r}^{b_2} \geq C_2\delta_T^{3/2}\underline{f}_T > \zeta_{T,d}.
\end{align}
We will now show that $\left|b_2 - r_2\right|\left(\Delta_2^{q_2}\right)^{2/3} \leq C_3\left(\log T\right)^{1/3}$. Following the same process as in Step 3.1 and assuming now that $b_2 < r_2$, we have that for $b^* \in \left\lbrace c_{k^*}^r,\ldots,c_{k^*_2}^r - 1\right\rbrace$, 
\begin{equation}
\label{r2_relationship_contradiction_trend}
\left(C_{c_{k^*}^r,c_{k^*_2}^r}^{r_2}(\boldsymbol{X_{q_2}})\right)^2 > \left(C_{c_{k^*}^r,c_{k^*_2}^r}^{b^*}(\boldsymbol{X_{q_2}})\right)^2
\end{equation}
is implied by $\min \left\lbrace \left|b^* - r_2\right|,c_{k_2^*}^r - r_2 \right\rbrace > C_3\left(\log\left(Td^{1/4}\right)\right)^{1/3}/\left(\Delta_2^{q_2}\right)^{2/3}$. However, following the same procedure as in Step 3.1 we can show that for sufficiently large $Td^{1/4}$, $$\min \left\lbrace c_{k_2^*}^r - r_2, r_2 - c_{k^*}^r \right\rbrace > C_3\frac{\left(\log\left(Td^{1/4}\right)\right)^{1/3}}{\left(\Delta_2^{q_2}\right)^{2/3}}.$$ Thus, \eqref{r2_relationship_contradiction_trend} is implied by $\left|b^* - r_2\right|\left(\Delta_2^{q_2}\right)^{2/3} > C_3(\log\left(Td^{1/4}\right))^{1/3}$. Therefore, $\left|b_2 - r_2\right|\left(\Delta_2^{q_2}\right)^{2/3} > C_3 \left(\log \left(Td^{1/4}\right)\right)^{1/3}$ would necessarily mean that $G_{c_{k^*}^r,c_{k^*_2}^r}^{r_2} > G_{c_{k^*}^r,c_{k^*_2}^r}^{b_2}$, which is not true by the definition of $b_2$. Having said this, we conclude that $\left|b_2 - r_2\right|\left(\Delta_2^{q_2}\right)^{2/3} \leq C_3\left(\log\left(Td^{1/4}\right)\right)^{1/3}$.

Having detected $r_2$, then our algorithm will proceed in the interval $[s,e]=[c_{k^*_2}^r, T]$ and all the change-points will get detected one by one since Step 3.2 will be applicable as long as there are previously undetected change-points in $[s,e]$. Denoting by $\hat{r}_j$ the estimation of $r_j$ as we did in the statement of the theorem and for $[s^*,e^*]$ being the interval where the isolation and detection of $r_j$ occurs (in the way that we explained in Steps 3.1 and 3.2) allow us to denote by $q_j:= {\rm argmax}_{m=1,\ldots,d}\left\lbrace C_{s^*, e^*}^{\hat{r}_j}(\boldsymbol{X_m})\right\rbrace$. Then, Steps 3.1 and 3.2 have shown that MID will detect all change-points one by one and $\left|\hat{r}_j - r_j\right|\left(\Delta_j^{q_j}\right)^{2/3} \leq C_3\left(\log\left(Td^{\frac{1}{4}}\right)\right)^{1/3}, \quad \forall j \in \left\lbrace 1,\ldots,N \right\rbrace.$
\vspace{0.1in}
\\
{\textbf{Step 4:}} The arguments given in Steps 1-3 hold in $\tilde{A}_T^* \cap \tilde{B}_T$.  At the beginning of the algorithm, $s=1, e=T$ and for $N\geq 1$, there exist $k_1\in \left\lbrace 1,\ldots, K \right\rbrace$ such that $s_{k_1} = s, e_{k_1} \in I_1^R$ and $k_2\in \left\lbrace 1,\ldots, K \right\rbrace$ such that $s_{k_2} \in I_N^L, e_{k_2} = e$, where $I_j^R$ and $I_j^L$ are defined in \eqref{isolating_intervals} of the main paper. As in our previous steps, w.l.o.g. assume that $r_1 \leq T- r_N + 1$, meaning that $r_1$ gets isolated and detected first in an interval $[s,c_{k^*}^r]$, where $c_{k^*}^r \leq e_{k_1}$. Then, $\hat{r}_1 = {\rm argmax}_{s \leq t < c_{k^*}^r}G_{s,c_{k^*}^r}^t$ is the estimated location for $r_1$ and $\left|r_1-\hat{r}_1\right|\left(\Delta_1^{q_1}\right)^{2/3}\leq C_3\left(\log \left(Td^{1/4}\right)\right)^{1/3}$. After this, the algorithm continues in $[c_{k^*}^r,T]$ and keeps detecting all the change-points as explained in Step 3. It is important to note that there will not be any double detection issues because naturally, at each step of the algorithm, the new interval $[s,e]$ does not include any previously detected change-points.

Once all the change-points have been detected one by one, then $[s,e]$ will have no other change-points in it. Our method will keep interchangeably checking for possible change-points in right- and left-expanding intervals denoted by $[s^*,e^*]$. MID will not detect anything in $[s^*,e^*]$ since $\forall b \in [s^*,e^*)$, using \eqref{set_A_T_linear} we have that
\begin{equation}
\nonumber G_{s^*,e^*}^{b} \leq H_{s^*,e^*}^{b} + \sqrt{8\log\left(Td^{\frac{1}{4}}\right)} = \sqrt{8\log\left(Td^{1/4}\right)} < C_1\sqrt{\log \left(Td^{1/4}\right)} \leq \zeta_{T,d}.
\end{equation}
After not detecting anything in all intervals of the above form, then the algorithm concludes that there are not any change-points in $[s,e]$ and stops.
\end{proof}



\end{document}